\documentclass[useAMS,usenatbib]{mn2e}
\usepackage{hyperref}
\usepackage[utf8]{inputenc}
\usepackage{graphicx,url,times,subfig}
\newif\ifpdf
  \ifx\pdfoutput\undefined
    \pdffalse
  \else
    \ifnum\pdfoutput=1
      \pdftrue
    \else
      \pdffalse
    \fi
  \fi
\ifpdf\hypersetup{
pdftitle={SubHaloes going Notts: The SubHalo-Finder Comparison Project},
pdfauthor={Julian Onions},
pdfkeywords={N-body simulations, haloes evolution, dark matter},
}
\fi

\newcommand{\ahf}{\textsc{ahf}}
\newcommand{\subfind}{\textsc{subfind}}

\newcommand{\gadget}{\textsc{gadget3}}
\newcommand{\voboz}{\textsc{voboz}}
\newcommand{\mendieta}{\textsc{mendieta}}

\newcommand{\adaptahop}{\textsc{adaptahop}}

\newcommand{\htd}{\textsc{hot3d}}
\newcommand{\hsd}{\textsc{hot6d}}

\newcommand{\rockstar}{\textsc{rockstar}}
\newcommand{\hbt}{\textsc{hbt}}
\newcommand{\hsf}{\textsc{hsf}}
\newcommand{\stf}{\textsc{stf}}
\newcommand{\hkpc}{{\ifmmode{h^{-1}{\rm kpc}}\else{$h^{-1}$kpc}\fi}}

\newcommand{\Fig}[1]{Figure~\ref{#1}}
\def\vmax{$v_{\rm max}$}
\def\Rmax{$r_{\rm max}$}
\newcommand{\rth}{$R_{200}$}
\newcommand{\subhalos}{subhaloes}
\renewcommand{\S}{Section~}

\newlength{\figwidth}
\newlength{\resplot}
\title[SubHalo-Finder Comparison]
{SubHaloes going Notts: The SubHalo-Finder Comparison Project}
\author[Onions et al.]
{Julian~Onions,$^1$\thanks{E-mail: \href{mailto:julian.onions@gmail.com}{julian.onions@gmail.com}}
  Alexander~Knebe,$^2$
  Frazer~R.~Pearce,$^1$ 
  Stuart~I.~Muldrew,$^1$ 
\newauthor
  Hanni~Lux,$^1$
  Steffen~R.~Knollmann,$^2$ 
  Yago Ascasibar,$^2$
  Peter Behroozi,$^{3,4,5}$
\newauthor
  Pascal Elahi,$^{6}$
  Jiaxin Han,$^{6,7,8}$
  Michal Maciejewski,$^{9}$
  Manuel E. Merch\'{a}n,$^{10}$
\newauthor
  Mark Neyrinck,$^{11}$
  Andr\'{e}s N. Ruiz,$^{10}$ 
  Mario A. Sgr\'{o},$^{10}$ 
  Volker Springel,$^{12,13}$
\newauthor
 and Dylan Tweed$^{14}$
\\
  $^1$School of Physics \& Astronomy, University of Nottingham, Nottingham, NG7 2RD, UK\\
  $^2$Departamento de F\'{ı}sica Te\'{o}rica, M\'{o}dulo C-15, Facultad de Ciencias, 
  Universidad Aut\'{o}noma de Madrid, 28049 Cantoblanco, Madrid, Spain\\
  $^{3}$Kavli Institute for Particle Astrophysics and Cosmology, Stanford, CA 94309, USA\\
  $^{4}$Physics Department, Stanford University, Stanford, CA 94305, USA\\
  $^{5}$SLAC National Accelerator Laboratory, Menlo Park, CA 94025, USA\\
  $^{6}$Key Laboratory for Research in Galaxies and Cosmology, Shanghai Astronomical Observatory, Shanghai 200030, China\\
  $^{7}$Graduate School of the Chinese Academy of Sciences, 19A, Yuquan Road, Beijing, China \\
  $^{8}$Institute for Computational Cosmology, Department of Physics, Durham University, South Road, Durham DH1 3LE, UK\\
  $^{9}$Max-Planck-Institut f\"{u}r Astrophysik, Garching,
  Karl-Schwarzschild-Stra\ss e 1, 85741 Garching bei M\"{u}nchen,
  Germany\\
  $^{10}$Instituto de Astronom\'{ı}a Te\'{o}rica y Experimental (CCT C\'{o}rdoba, CONICET, UNC), Laprida 922, X5000BGT, C\'{o}rdoba, Argentina\\
  $^{11}$Department of Physics and Astronomy, Johns Hopkins University, 
  3701 San Martin Drive, Baltimore, MD 21218, USA\\
  $^{12}$Heidelberg Institute for Theoretical Studies, Schloss-Wolfsbrunnenweg 35, 69118 Heidelberg, Germany\\ 
  $^{13}$Zentrum f\"ur Astronomie der Universit\"at Heidelberg, ARI, M\"{o}nchhofstr. 12-14, 69120 Heidelberg, Germany \\
  $^{14}$Institut d'Astrophysique Spatiale, CNRS/Universite Paris-Sud 11, 91405 Orsay, France \\
}

\begin{document}

\pagerange{\pageref{firstpage}--\pageref{lastpage}} \pubyear{2011}\volume{0000}

\maketitle

\label{firstpage}

\begin{abstract}
  We present a detailed comparison of the substructure properties of a
  single Milky Way sized dark matter halo from the Aquarius suite
  (Springel et al.) at five different resolutions, as identified by a
  variety of different (sub-)halo finders for simulations of cosmic
  structure formation. These finders span a wide range of techniques
  and methodologies to extract and quantify substructures within a
  larger non-homogeneous background density (e.g. a host halo). This
  includes real-space, phase-space, velocity-space and time-space based finders, as
  well as finders employing a Voronoi tessellation, friends-of-friends
  techniques, or refined meshes as the starting point for locating
  substructure.
  A common post-processing pipeline was used to uniformly analyse the
  particle lists provided by each finder.  We extract quantitative and
  comparable measures for the \subhalos, primarily focusing on mass
  and the peak of the rotation curve for this particular study. We
  find that all of the finders agree extremely well on the presence
  and location of substructure and even for properties relating to the
  inner part part of the subhalo (e.g. the maximum value of the
  rotation curve). For properties that rely on particles near the
  outer edge of the subhalo the agreement is at around the 20 per cent
  level. We find that basic properties (mass, maximum circular
  velocity) of a subhalo can be reliably recovered if the subhalo
  contains more than 100 particles although its presence can be
  reliably inferred for a lower particle number limit of 20.
  We finally note that the logarithmic slope of the subhalo cumulative
  number count is remarkably consistent and $<1$ for all the finders
  that reached high resolution. If correct, this would indicate that the larger and
  more massive, respectively, substructures are the most dynamically
  interesting and that higher levels of the (sub-)subhalo hierarchy
  become progressively less important.
\vspace{2cm}
\end{abstract}


\section{Introduction} \label{sec:introduction}

The growth of structure via a hierarchical series of mergers is now a
well established paradigm \citep{white_1978}. As larger structures
grow they subsume small infalling objects. However the memory of the
existence of these substructures is not immediately erased, either in
the observable Universe (where thousands of individual galaxies within
a galaxy cluster are obvious markers of this pre-existing structure)
or within numerical models, first noted for the latter by
\citet{klypin_overmerging_1999}.

Knowing the properties of substructure created in cosmological
$N$-body simulations allows the most direct comparison between these
simulations and observations of the Universe. The fraction of material
that remains undispersed and so survives as separate structures within
larger haloes is an important quantity for both studies of dark-matter
detection \citep{springel_prospects_2008,kuhlen2008dark,vogelsberger2009phase,zavala_2010} 
and the apparent overabundance of
substructure within numerical models when compared to observations
\citep{klypin_1999,moore_1999}. The mass and radial position of the
most massive Milky Way satellites seem to raise new concerns for our
standard $\Lambda$CDM cosmology
\citep{boylan_toobig_2011,boylan_mwlcdm_2011,dicintio_2011,ferrero_2011},
while differences between the simulated and observed internal density
profiles of the satellites seems to have been reconciled by taking
baryonic effects into account \citep[e.g.][]{Oh_2011,pontzen_2011}.
We are certain that between 5 per cent and 10 per cent of the material
within simulated galactic sized haloes exists within bound
substructures \citep[e.g.][]{gao_2004,DeLucia_2004,contini_2011} and a substantial part of
the host halo has formed from disrupted subhalo material
\citep[e.g.][]{gill_2004b,knebe_2005,warnick_2008,cooper_2010,libeskind_2011}.

Quantification of the amount of substructure (both observationally and
in simulations of structure formation) is therefore an essential tool
to what is nowadays referred to as ``Near-Field Cosmology''
\citep{freeman_2002} and attempts to do so in numerical models have
followed two broad approaches: either a small number of individual
haloes are simulated at exquisite resolution \citep[e.g.][respectively
the Via Lactea, Aquarius and GHalo projects]{diemand_vl2_2008,
  springel_aquarius_2008, stadel_ghalo_2009} or a larger
representative sample of the Universe is modelled in order to quantify
halo-to-halo substructure variations \citep[e.g.][who used the
Millennium simulation, Millennium II simulation and the Bolshoi
simulation, respectively]{angulo_2009, boylan_mill2_2009,
  klypin_bolshoi_2011}.  As this paper studies the convergence of
halo-finders within a single halo we can add nothing to the topic of
halo-to-halo substructure variations.

In a very
comprehensive study that included 6 different haloes and 5 levels of
resolution \citet{springel_aquarius_2008} utilised their substructure
finder \subfind\ to detect around 300000 substructures within the
virial radius of their best resolved halo. They found that the number
counts of substructures per logarithmic decade in mass falls with a
power law index of at most 0.93, indicating that smaller substructures
are progressively less dynamically important and that the central
regions of the host dark matter halo are likely to be dominated by a
diffuse dark matter component composed of hundreds of thousands of
streams of tidally stripped material. \citet{maciejewski_2011}
confirmed the existence and properties of this stripped material using
a 6-dimensional phase-space finder \hsf. A similar power law index was
also found for the larger cosmological studies
\citep{boylan_mill2_2009, angulo_2009}. For the Bolshoi simulation
\citet{klypin_bolshoi_2011} find results that are in agreement with
their re-analysis of the Via Lactea II of \citet{diemand_vl2_2008}
with an abundance of \subhalos\ falling as the cube of the subhalo
rotation velocity. Rather than the present value of the maximum
rotation velocity they prefer to use the value that the subhalo had
when it first became a subhalo (i.e. on infall). This negates the
effects of tidal stripping and harassment within the cluster
environment but makes it difficult for us to directly compare as
we have generally only used the final $z=0$ snapshot for this
comparison study.

In recent years there has not only been a number of different groups
performing billion particle single-halo calculations, there has also
been an explosion in the number of methods available for quantifying
the size and location of the structures within such an $N$-body
simulation \citep[see, for instance, Fig.1
in][]{knebe_haloes_2011}. In this paper we extend the halo finder
comparison study of \citet{knebe_haloes_2011} to examine how well
these finders extract the properties of those haloes that survive the
merging process and live within larger haloes. While this issue has
already been addressed by \citet{knebe_haloes_2011} it was
nevertheless only in an academic way where controlled set-ups of
individual \subhalos\ placed into generic host haloes were studied;
here we apply the comparison to a fully self-consistently formed dark
matter halo extracted from a cosmological simulation. As the results
of credible and reliable subhalo identification have such important
implications across a wide range of astrophysics it is essential to
ask how well the (sub-)halo finders perform at reliably extracting 
\subhalos. This still leaves open the question of how well different
modern gravity solvers compare when performing the same simulation but
at least we can hope to ascertain whether or not -- given the same set
of simulation data -- the different finders will arrive at the same
conclusions about the enclosed subhalo properties. We intend this
paper to form the first part of a series of comparisons. It primarily
focuses on the most relevant subhalo properties, i.e. location, mass
spectrum and the distribution of the value of the peak of the
\subhalos' rotation curve.

In \S\ref{sec:Finders} we begin by summarising the eleven substructure
finders that have participated in this study, focusing upon any
elements that are of particular relevance for substructure finding. In
\S\ref{sec:Data} we introduce the Aquarius dataset that the described
finders analysed for this study. Both a qualitative and a quantitative
comparison between the finders is contained in \S\ref{sec:Comparison}
which also contains a discussion of our results, before we summarise
and conclude in \S\ref{sec:Summary}.

\begin{figure*}
 \includegraphics[width=\figwidth]{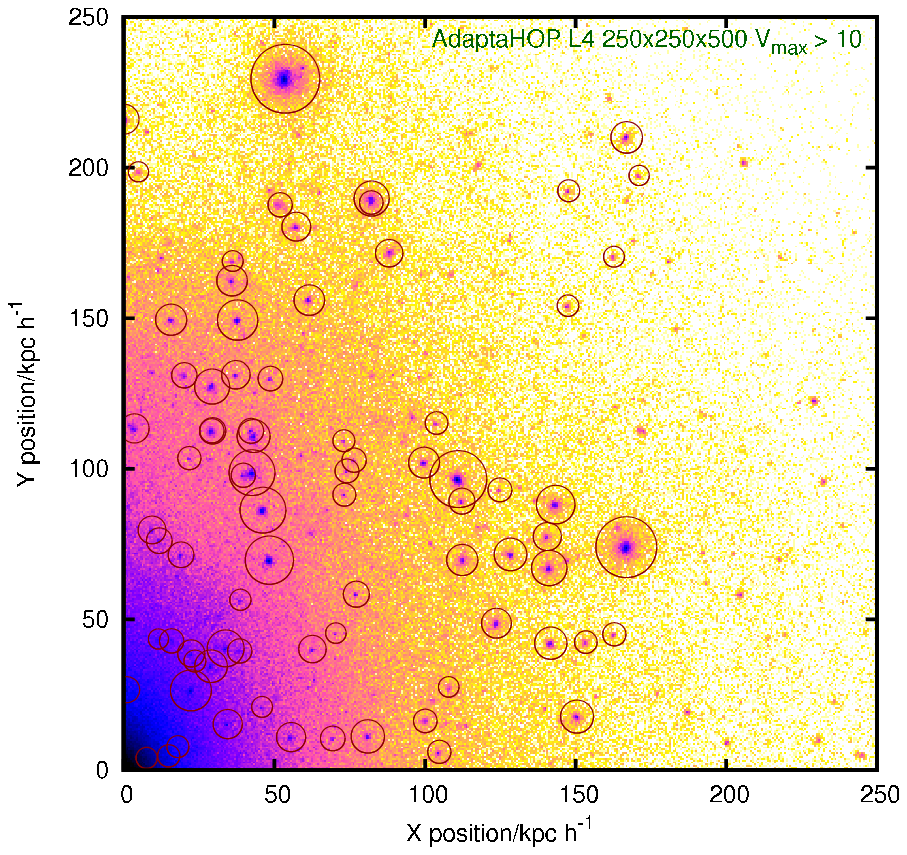}
 \includegraphics[width=\figwidth]{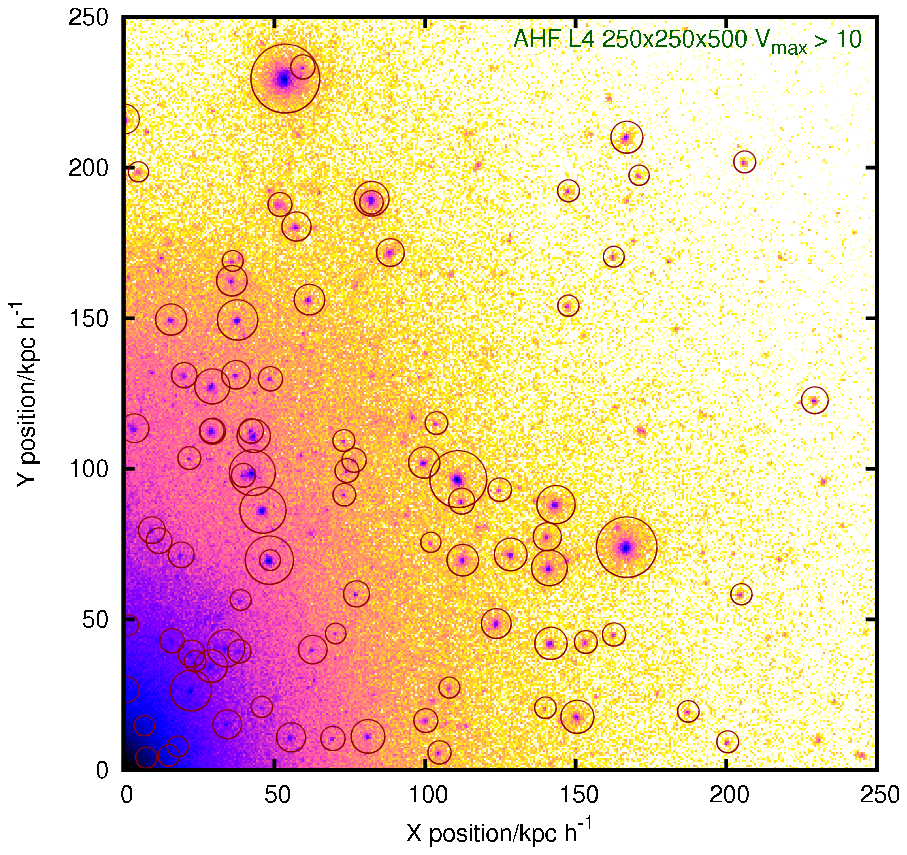} \\
 \includegraphics[width=\figwidth]{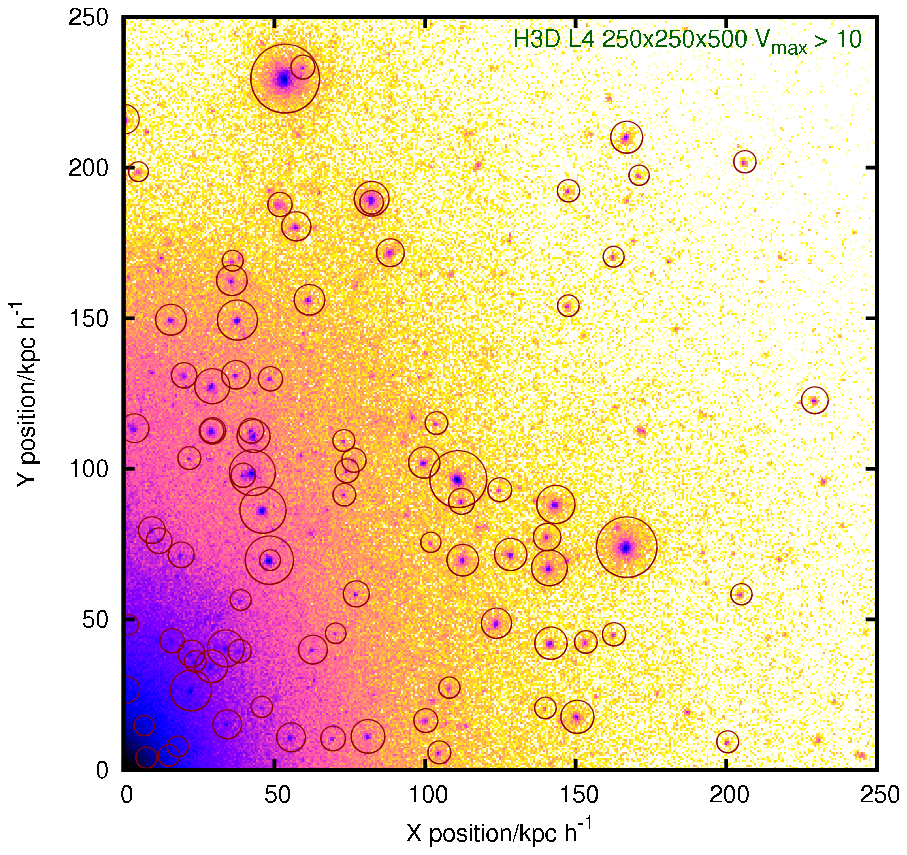}
 \includegraphics[width=\figwidth]{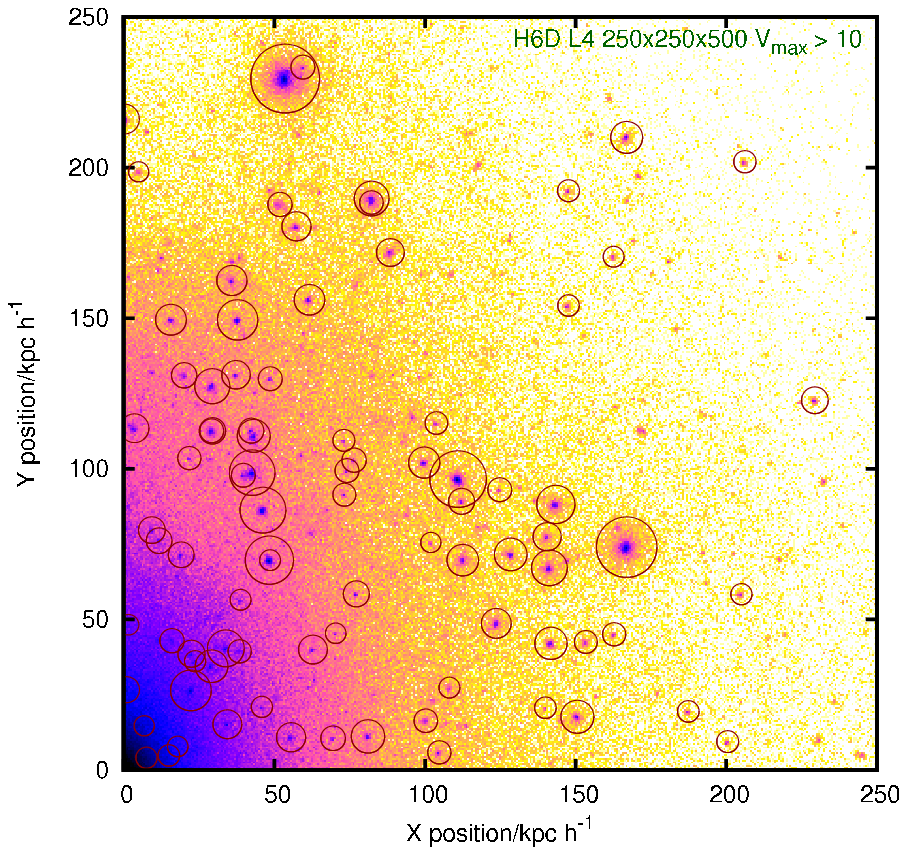} \\
 \includegraphics[width=\figwidth]{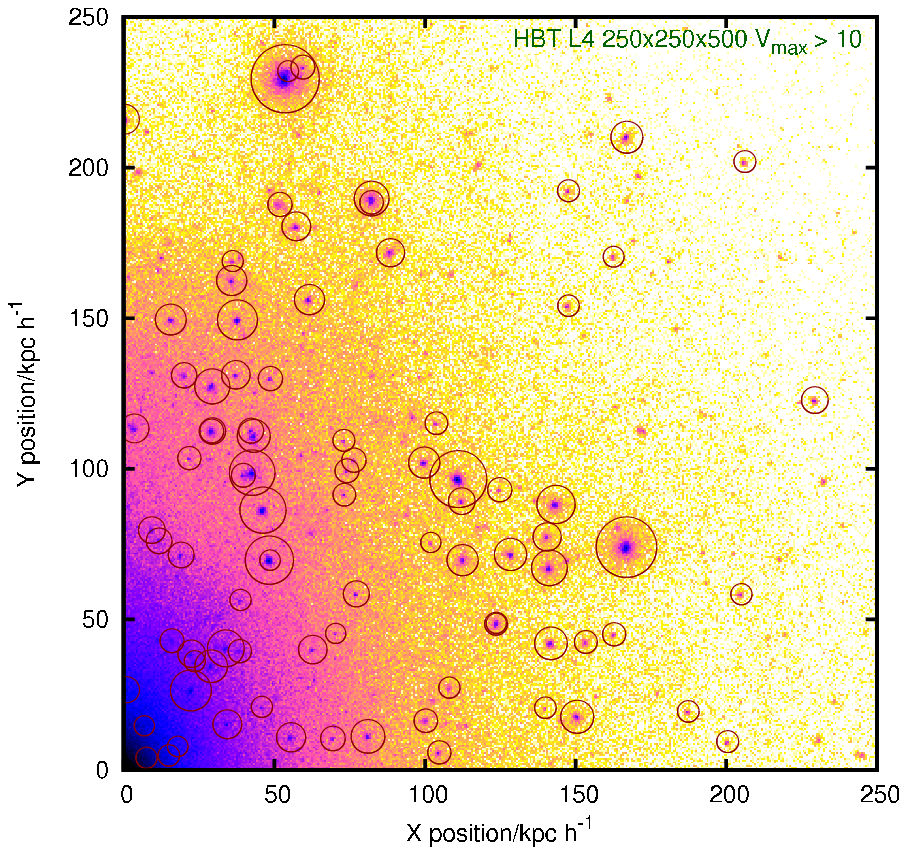} 
 \includegraphics[width=\figwidth]{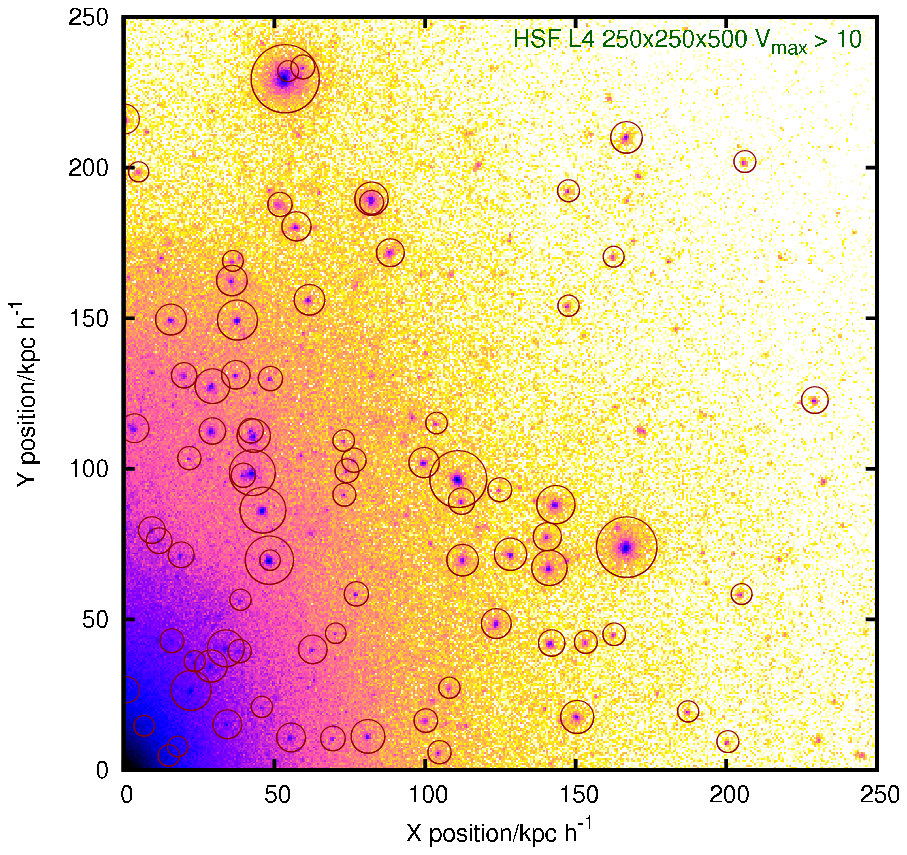}
\caption{The images show the smoothed dark matter density within an
  quadrant at resolution level 4. In each panel the overplotted circles indicate
  the location of the recovered \subhalos\ for the finder labelled at
  the top of each panel. They are scaled proportionally using
  \vmax. Only \subhalos\ with a \vmax\ greater than 10 km/s are
  shown.}
\label{fig:haloes}
\end{figure*}

\begin{figure*}
\ContinuedFloat
 \includegraphics[width=\figwidth]{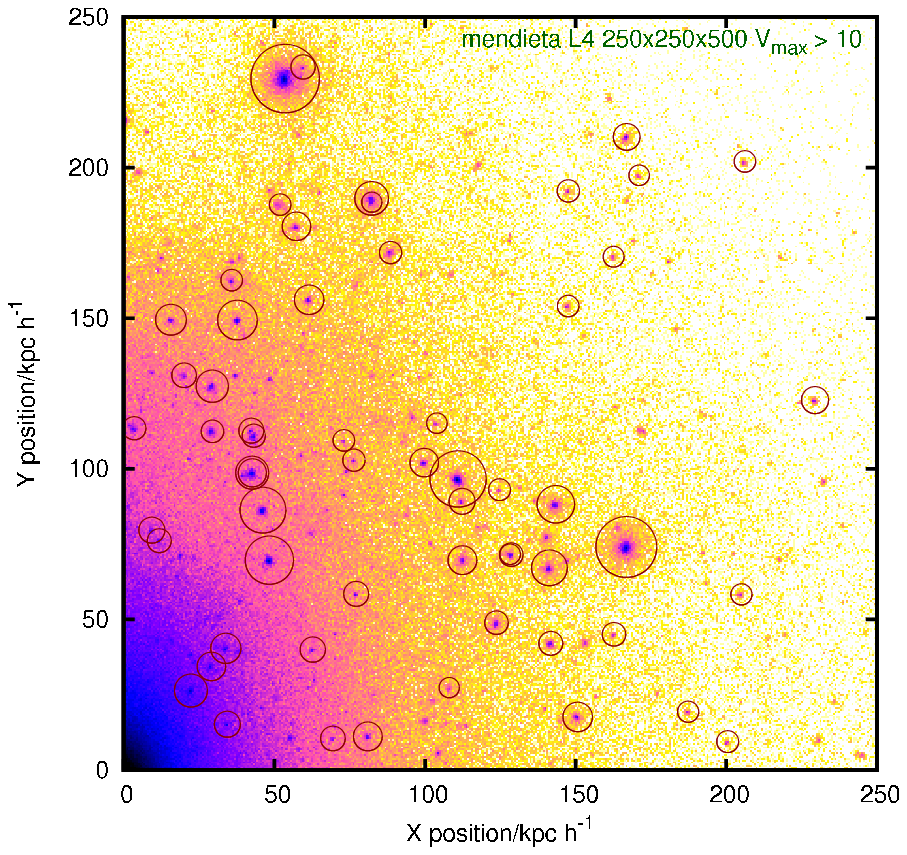} 
 \includegraphics[width=\figwidth]{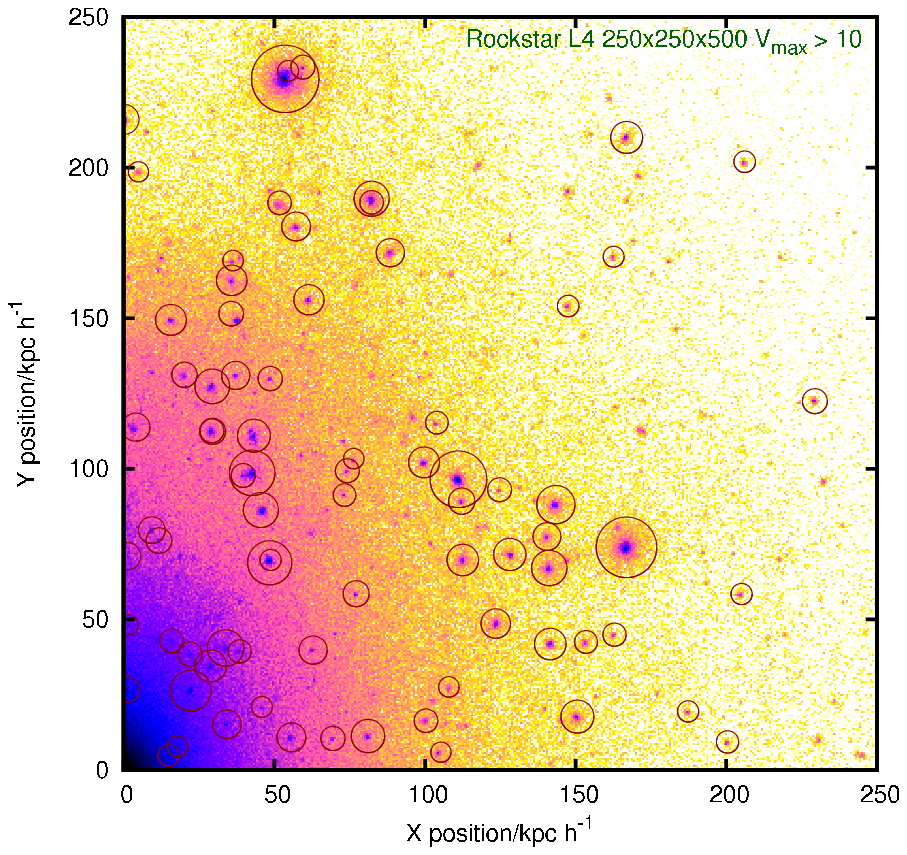} \\
 \includegraphics[width=\figwidth]{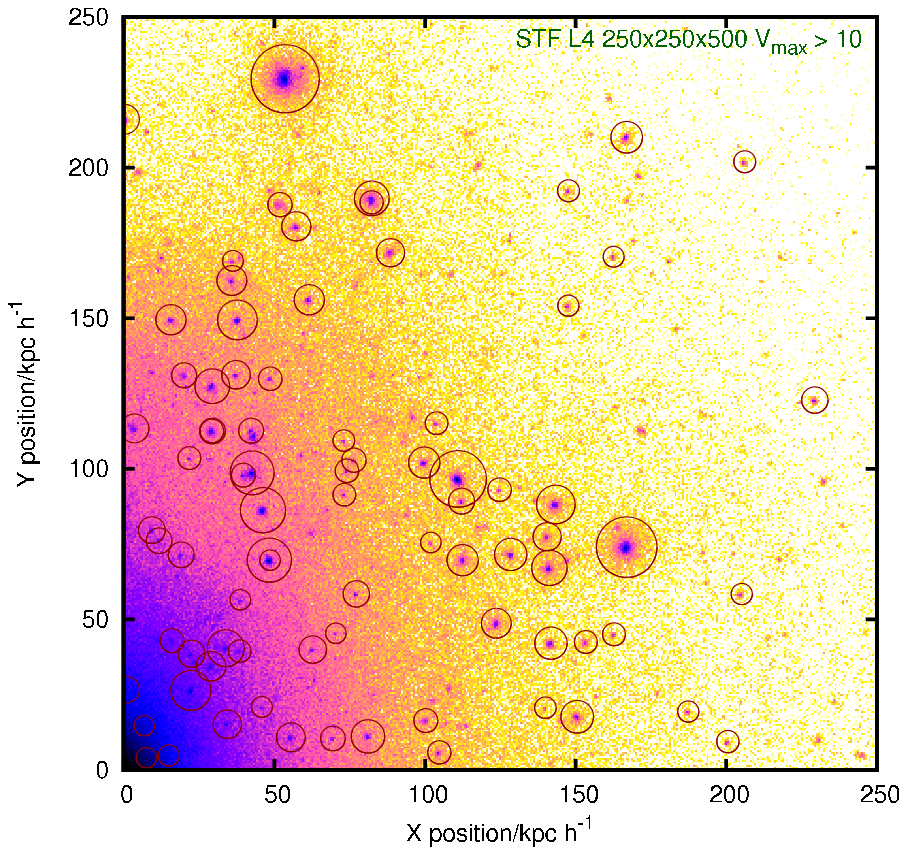} 
 \includegraphics[width=\figwidth]{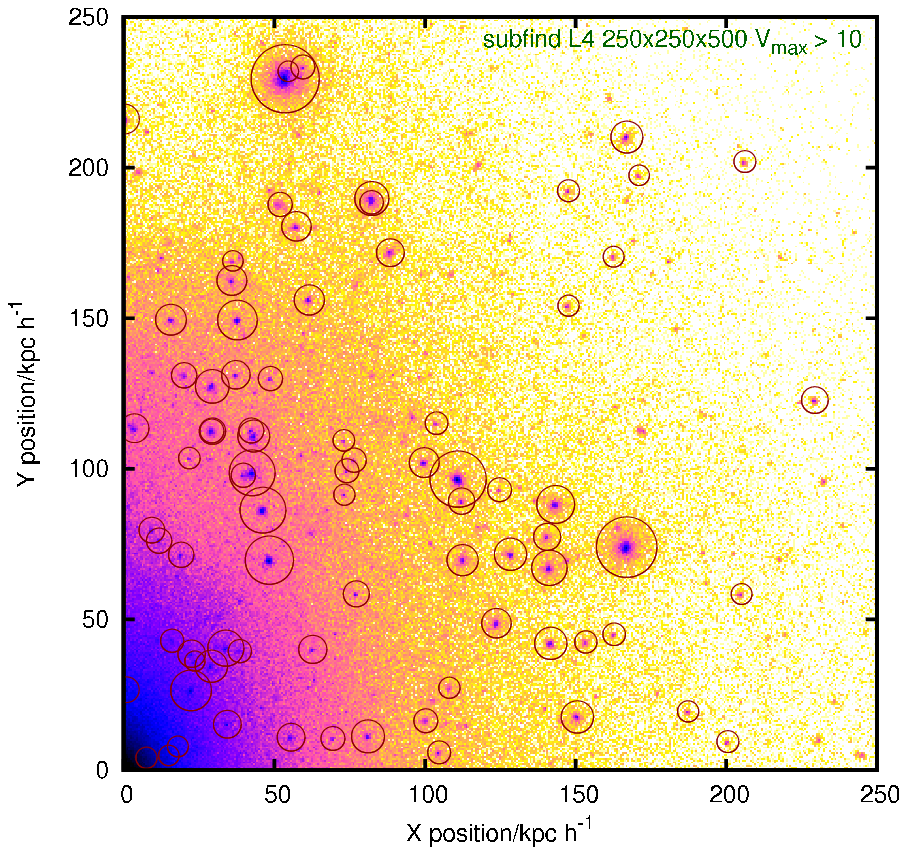} \\
 \includegraphics[width=\figwidth]{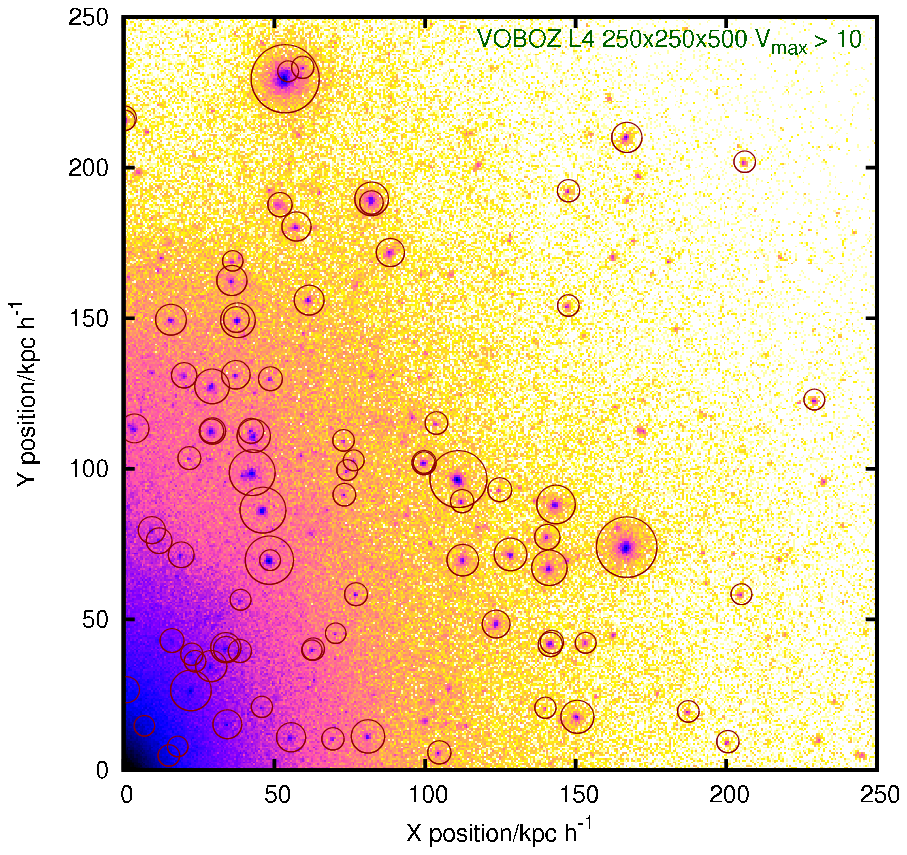} 
\caption[]{(continued) Recovered subhalo locations and \vmax\ scale by
  labelled finder.}
\label{fig:haloes2}
\end{figure*}

\section{The SubHalo Finders} \label{sec:Finders}

In this section we present the (sub-)halo finders participating in the
comparison project in alphabetical order. Please note that we
primarily only provide references to the actual code description
papers and not an exhaustive portrait of each finder as this would be
far beyond the scope of this paper. While the general mode of
operation can be found elsewhere, we nevertheless focus here on the
way each code collects and defines the set of particles belonging to a
subhalo: as already mentioned before, those particle lists are
subjected to a common post-processing pipeline and hence the retrieval
of this list is the only relevant piece of information as far as the
comparison in this particular paper is concerned.

\subsection{ADAPTAHOP (Tweed)}
\adaptahop\ is a full topological algorithm. The first stage consist
in estimating a local density using a 20 particles SPH
kernel. Particles are then sorted into groups around a local density
maximum. And saddle points act as links between group. All groups are
first supposed to be one single entity, that we hierarchically divide
into smaller groups, by using an increasing density threshold. Haloes
are then defined as groups of groups linked by saddle points
corresponding to densities higher that 80 times the mean DM
density. By increasing the threshold we further detail the structure
of the halo as a node structure tree. Where a node is either a local
maxima, or a group of particules connecting higher level nodes.  After
using this bottom to top approach, the (sub)haloes are defined using a
top to bottom approach, hierarchically regrouping nodes so that a
sub(sub)halo has a smaller mass than its host (sub)halo. Each particle
belongs to a single structure either a halo or a subhalo.  The
(sub)haloes' centers are defined as the position of its particles with
the highest SPH density. We need to stress that no unbinding
procedures are used in this algorithm, at the risk of overestimating
the number/misidentification of \subhalos\ with a low number of
particles.

\subsection{AHF (Knollmann \& Knebe)}
The halo finder \ahf\footnote{\ahf\ is freely
  available from
  \href{http://www.popia.ft.uam.es/AMIGA}{http://www.popia.ft.uam.es/AMIGA}}
(\textsc{amiga} Halo Finder) is a spherical overdensity finder that
identifies (isolated and sub-)haloes as described in
\citet{gill_evolution_2004} as well as
\citet{knollmann_ahf:_2009}. The initial particle lists are obtained
by a rather elaborate scheme: for each subhalo the distance to its
nearest more massive (sub-)halo is calculated and all particles within
a sphere of radius half this distance are considered prospective
subhalo constituents. This list is then pruned by an iterative
unbinding procedure using the (fixed) subhalo centre as given by the
local density peak determined from an adaptive mesh refinement
hierarchy. For more details we refer the reader to aforementioned code
description papers as well as the online documentation.

\subsection{Hierarchical Bound-Tracing (HBT) (Han)}
\hbt\ \citep{han_resolving_2011} is a tracing algorithm working in the
time domain of each \subhalos' evolution. Haloes are identified with a
Friends-of-Friends algorithm and halo merger trees are
constructed. \hbt\ then traverses the halo merger trees from the
earliest to the latest time and identifies a self-bound remnant for
every halo at every snapshot after infall.  Care has been taken to
ensure that \subhalos\ are robustly traced over long periods.  The
merging hierarchy of progenitor haloes are recorded to efficiently
allow satellite-satellite mergers or satellite accretion.\footnote{It
  should be noted that \hbt\ had access to the full snapshot data for
  Aquarius-A.}

\subsection{HOT+FiEstAS (\htd\ \& \hsd) (Ascasibar)}
HOT+FiEstAS is a general-purpose clustering analysis tool, still under
development, that performs the unsupervised classification of a
multidimensional data set by computing its Hierarchical Overdensity
Tree (HOT), analogous to the Minimal Spanning Tree (MST) in Euclidean
spaces, based on the density field returned by the Field Estimator for
Arbitrary Spaces \citep[FiEstAS][]{ascasibar_2005,ascasibar_2010}.
As explained in \citet{knebe_haloes_2011} in the context of halo
finding, HOT+FiEstAS identifies objects with density maxima, either
in configuration space (considering particle positions alone, \htd)
or in the full, six-dimensional phase-space of particle positions and
velocities (\hsd ). In both cases, the boundary of an object is always
set by the isodensity contour crossing a saddle point, and its centre
is defined as the density-weighted average of its constituent particles.

The main difference with respect to the version used in
\citet{knebe_haloes_2011} is that there is now a post-processing stage,
akin to a `hard' expectation-maximization that is specifically tailored
to the problem of halo finding, where:

\begin{enumerate}
\item \Rmax\ and \vmax\ are computed for every object in the catalog.
\item Objects with more than 10 particles within \Rmax\ are labelled as (sub)-halo candidates.
\item Particles are assigned to the candidate that contributes most to the phase-space density 
at their location, approximating each candidate by a \citet{Hernquist_1990} 
sphere with the appropriate values of \Rmax\ and \vmax.
\end{enumerate}

Candidates are only kept if they contain more than 5 particles within
\Rmax\ \emph{and} the density within that radius is higher than 100
times the critical density. Although a detailed discussion is obviously
beyond the scope of this work, it is interesting to comment that some
of the objects discarded by the latter criterion seem to be numerical
artefacts, but others are clearly associated to filaments, streams,
and other loose -- yet physical, sometimes even gravitationally bound
-- structures. Since they are certainly not individual dark matter
(sub)-haloes, they can be simply discarded for our present purposes.

\subsection{Hierarchical Structure Finder (HSF) (Maciejewski)}
The Hierarchical Structure Finder (HSF) identifies objects as connected
self-bound particle sets above some density threshold.  This method
has two steps. Each particle is first linked to a local dark matter
phase-space density maximum by following the gradient of a
particle-based estimate of the underlying dark matter phase-space
density field. The particle set attached to a given maximum defines a
candidate structure. In a second step, particles which are
gravitationally unbound to the structure are discarded until a fully
self-bound final object is obtained.  For more details see
\citet{maciejewski_phasespace_2009}.

\subsection{MENDIETA (Sgr\'{o}, Ruiz \& Merch\'an)}
The \mendieta\ finder is a Friends-of-Friends based finder that is
used to obtain a dark matter halo. This prospective host halo is
subsequently refined by looking at peaks of increasing density by
reducing the linking length. This approach decomposes the halo into
its substructure plus other minor overdensities. In a final pass pass
unbound particles are removed by checking their associated
energies. \mendieta\ is described more fully in
\citet{m._a._sgro_hierarchical_2010}.

\subsection{ROCKSTAR (Behroozi)}
\rockstar\ (Robust Overdensity Calculation using K-Space Topologically
Adaptive Refinement) is a phase-space halo finder designed to maximize
halo consistency across timesteps \citep{behroozi_2011}.  The
algorithm first selects particle groups with a 3D Friends-of-Friends
variant with a very large linking length ($b = 0.28$). For each main
FOF group, Rockstar builds a hierarchy of FOF subgroups in phase-space
by progressively and adaptively reducing the linking length, so that a
tunable fraction (70\%, for this analysis) of particles are captured
at each subgroup as compared to the immediate parent group.  When this
is complete, Rockstar converts FOF subgroups into seed haloes
beginning at the deepest level of the hierarchy. If a particular group
has multiple subgroups, then particles are assigned to the subgroups’
seed haloes based on their phase-space proximity. This process is
repeated at all levels of the hierarchy until all particles in the
base FOF group have been assigned to haloes. Unbinding is performed
using the full particle potentials; halo centres and velocities are
calculated in a small region close to the phase-space density maximum.

\subsection{STF (Elahi)}
The STructure Finder Hierarchical Structure Finder \citep[(STF)]{elahi_peaks_2011} identifies
objects by utilizing the fact that dynamically distinct substructures
in a halo will have a {\em local} velocity distribution that differs
significantly from the mean, {\em i.e.} smooth background halo. This
method consists of two main steps, identifying particles that appear
dynamically distinct and linking this outlier population using a
Friends-of-Friends-like approach. Since this approach is capable of
not only finding \subhalos, but tidal streams surrounding \subhalos\ as
well as tidal streams from completely disrupted \subhalos, we also
ensure that a group is self-bound. Particles which are gravitationally
unbound to a candidate subhalo are discarded until a fully self-bound
is obtained or the object consists of fewer than 20 particles, at which
point the group is removed entirely.

\subsection{Subfind (Springel)}
\subfind\ identifies substructures as locally overdense, gravitationally
bound groups of particles. Starting with a halo identified through the
Friends-of-Friends algorithm, a local density is estimated for each particle
with adaptive kernel estimation using a prescribed number of smoothing
neighbours. Starting from isolated density peaks, additional particles are added
in sequence of decreasing density. Whenever a saddle point in the global density
field is reached that connects two disjoint overdense regions, the smaller
structure is treated as a substructure candidate, followed by merging the two
regions. All substructure candidates are subjected to an iterative unbinding
procedure with a tree-based calculation of the potential. The \subfind\ algorithm
is discussed in full in \citet{subfind_2001}.

\subsection{VOBOZ (Neyrinck)}
\voboz\ \citep{neyrinck_voboz:_2005} was developed to have little
dependence on free parameters.  Density peaks are found using a
Voronoi tessellation, which gives an adaptive, parameter-free estimate
of each particle's density and set of neighbours.  Each particle is
joined to the peak that lies up the steepest density gradient from
that particle.  A halo associated with a high density peak (which is
defined as the \voboz\ centre of the halo) will typically contain
smaller density peaks.  The significance of a halo is judged according
to the ratio of its central density to a saddle point joining the halo
to a halo with a higher central density, comparing to a Poisson point
process. For this project, we impose a 4-$\sigma$ significance threshold on \subhalos.
Particles not gravitationally bound to each halo are
iteratively removed, by comparing their potential energies (measured
as sums over all other particles) to kinetic energies with respect to
the velocity centroid of the halo's core (i.e.\ the particles that
directly jump up density gradients to the peak).  In the unbinding
process, the least-bound particles are removed first; for each halo,
the boundedness threshold reduces by a factor of $\sqrt{2}$ at each
iteration, until it reaches its true value.

\begin{figure*}
  \includegraphics[width=\resplot]{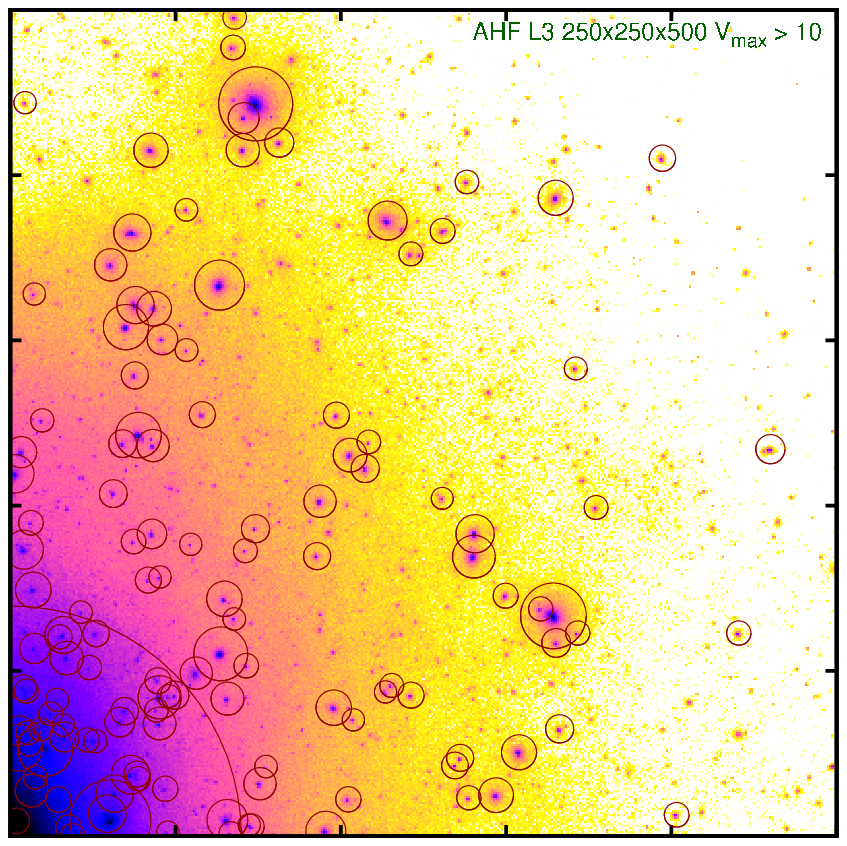} 
\includegraphics[width=\resplot]{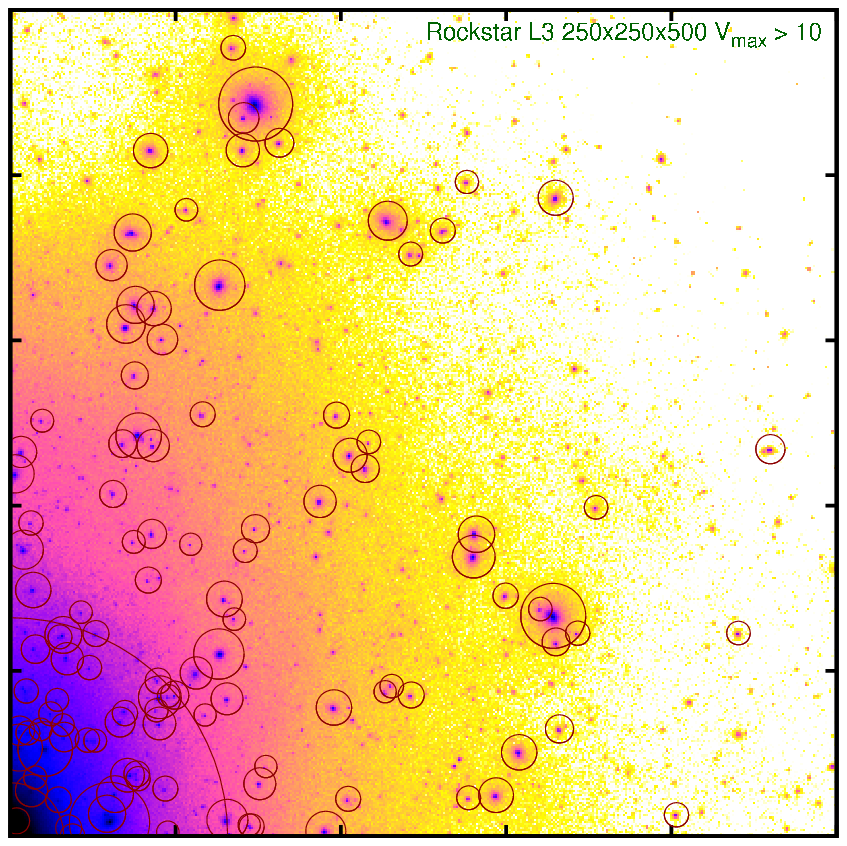}
  \includegraphics[width=\resplot]{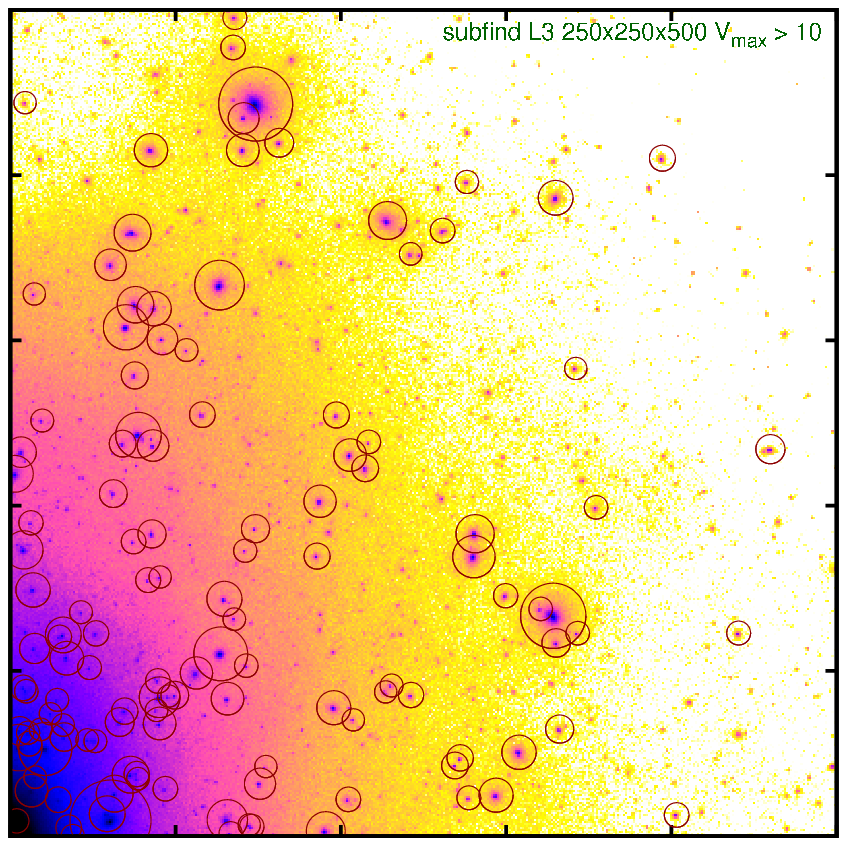}\\
  \includegraphics[width=\resplot]{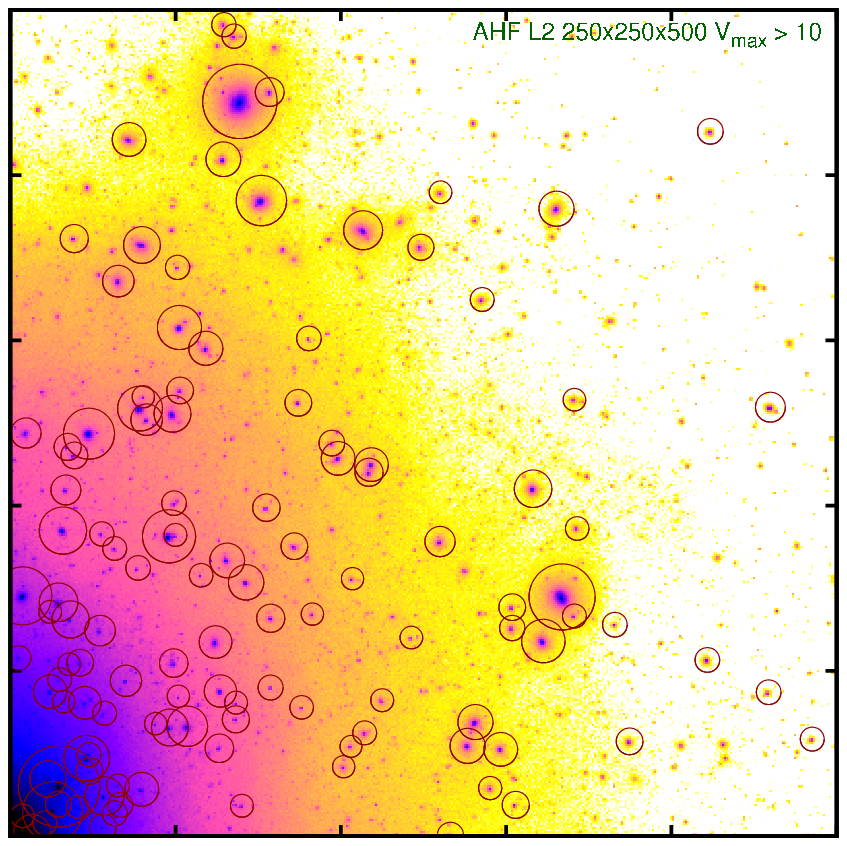} 
\includegraphics[width=\resplot]{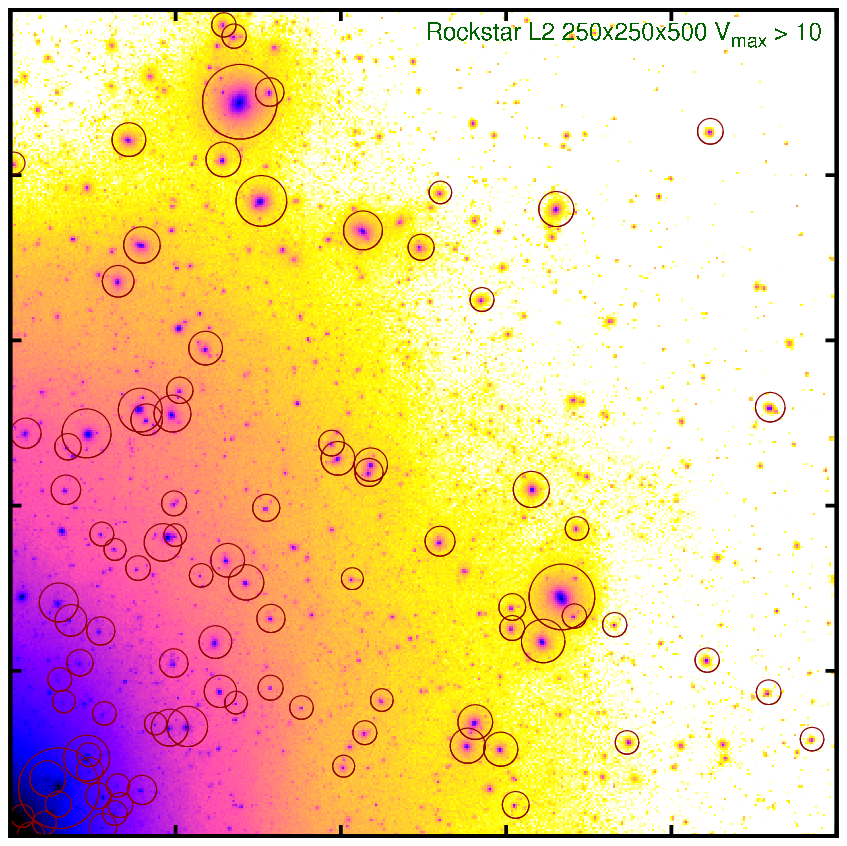}
  \includegraphics[width=\resplot]{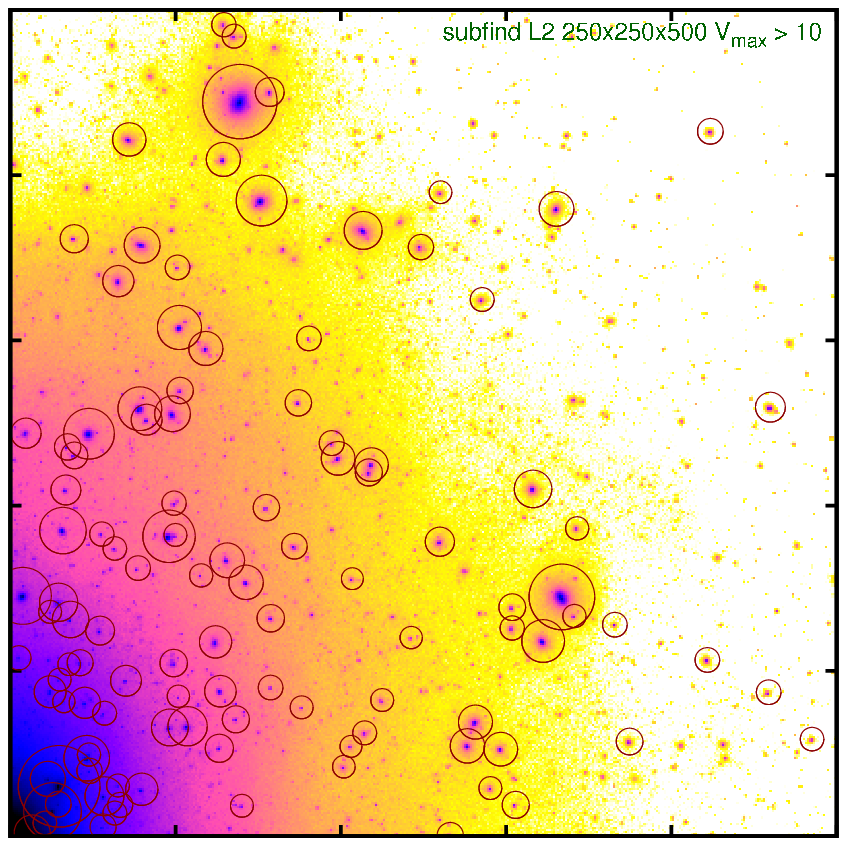}\\
  \includegraphics[width=\resplot]{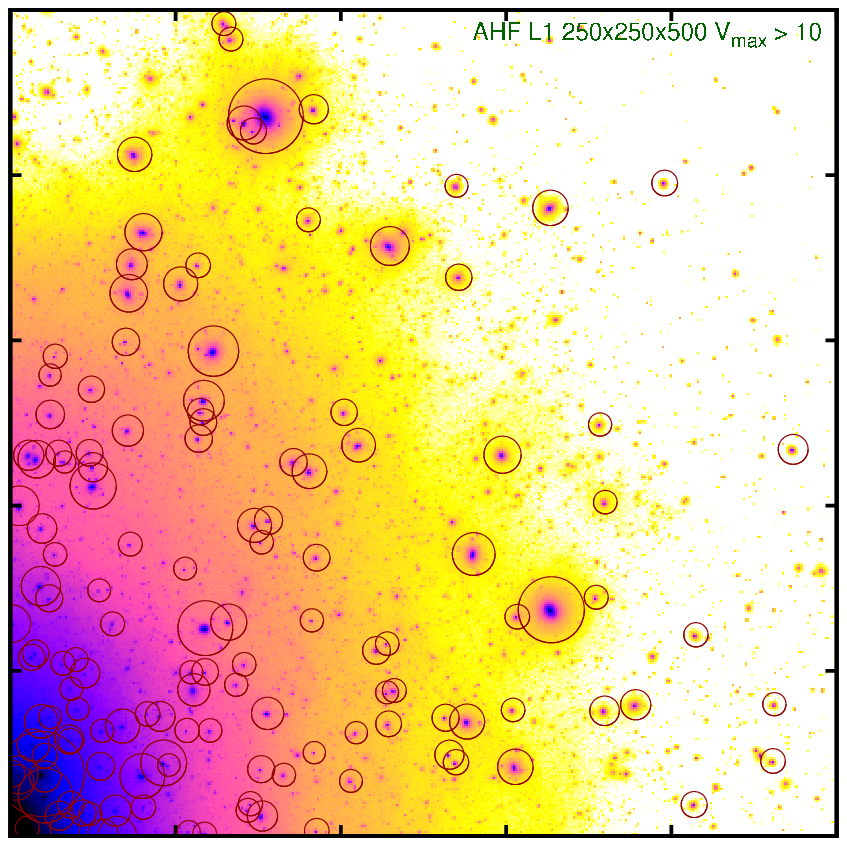} 
\includegraphics[width=\resplot]{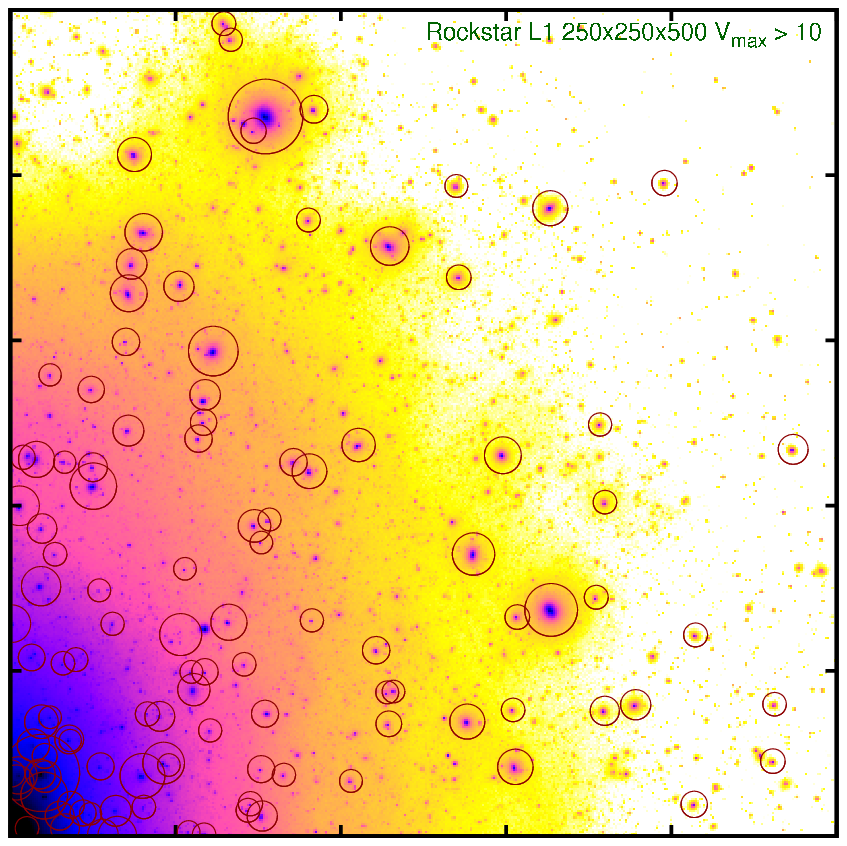}  
  \includegraphics[width=\resplot]{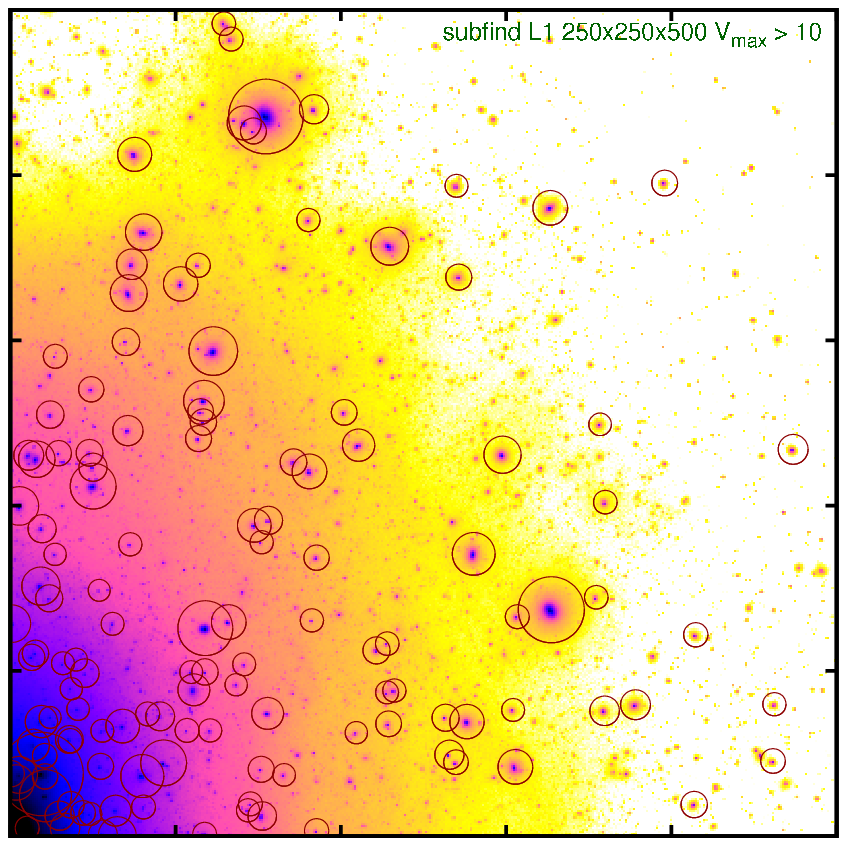}\\
\caption{Subhalo recovery as a function of resolution. Location and
  size of recovered substructure from level 3 to level 1 for the three
  finders that reached this level. In all panels \subhalos\ with
  \vmax $>10$ km/s are shown, scaled by \vmax\ as in \Fig{fig:haloes} and
  the background image is the smoothed dark matter density at that
  level. The relevant finder and level are labelled in the top right
  of each panel. The biggest change between levels is the additional
  small scale power moving the substructure locations.}
\label{fig:plotcomp}
\end{figure*}

\section{The Data} \label{sec:Data}

The data used for this paper forms part of the Aquarius project
\citep{springel_aquarius_2008}. It consists of multiple dark matter
only re-simulations of a Milky Way like halo at a variety of
resolutions performed using \gadget\ \citep[based on
\textsc{gadget2},][]{springel_cosmological_2005}.  We have used the
Aquarius-A halo dataset at $z=0$ for this project.  This provides 5
levels of resolution, varying in complexity from the 2.3 million
particles of the lowest resolution (i.e. level 5), up to the 4.25
billion particles of the highest resolution (i.e. level 1), as shown
in Table~\ref{tbl:Data}. The underlying cosmology for the Aquarius
simulations is the same as that used for the Millennium simulation
\citep{springel_millenium_2005} i.e. $\Omega_M = 0.25, \Omega_\Lambda
= 0.75, \sigma_8 = 0.9, n_s = 1, h = 0.73$. These parameters are
consistent with the latest WMAP data \citep{wmap_2011} although
$\sigma_8$ is a little high.  All the simulations were started at an
initial redshift of 127. Precise details on the set-up and performance
of these models can be found in \citet{springel_aquarius_2008}.

The participants were asked to run their subhalo finders on the
supplied data and to return a catalogue listing the substructures they
found.  Specifically they were asked to return a list of uniquely
identified substructures together with a list of all particles
associated with each subhalo.

Finders were initially run on the smallest dataset, the Aq-A-5 data.
This allowed for debugging of the common output format required by the
project and some basic checks on the internal consistency of the data
returned from each participant. Once this had been achieved each
participant scaled up to the higher resolution datasets, continuing
until they reached the limits of their finder and/or the computing
resources readily available to them. A summary of the number of
\subhalos\ found by each subhalo finder at the various levels is
contained in Table~\ref{tbl:ResultsCent} as well as the size of the
largest subhalo at level 4. All of the finders that participated in
this study completed the analysis of the level 4 dataset which is used
for the main comparison that follows and contains around 6 million
particles within the region considered, a sphere of radius 250 kpc/$h$
around a fiducial centre\footnote{We adopted a fixed and unique
  position for the host halo of $x=57060.4,y=52618.6,z=48704.8$
  kpc/$h$ independent of the resolution.}. Three of the finders (\ahf,
\rockstar\ \& \subfind) completed the analysis of the very
computationally demanding level 1 dataset. In addition to these \hbt\
and \hsf\ completed level 2 which contains around 160 million
particles within the region examined here.

\begin{table*}
  \caption{Summary of key numbers for each Aquarius level, the dataset
    used for this study. $N_h$ is the number of particles with the
    highest resolution (lowest individual mass).  $N_l$ the number of
    low resolution particles - the sum of the remainder.  $N_{250}$ is
    the number of high resolution particles found within a sphere of
    radius 250 kpc/h from the fiducial centre at each resolution ({\it
    i.e.} those of interest for this study). $M_p$ is the mass of one
    of these particles (in $\mathrm{M_{\sun}/h}$). $S$ is the resolution increase
    (mass decrease) for each level relative to level 5, and $S_p$ is
    the resolution increase relative to the previous level.  All
    particles are dark matter particles.}

  \label{tbl:Data}
  \begin{tabular}{l r r r r r r }
    \hline
    Data & $N_h$ & $N_l$ & $N_{250}$ & $M_p$ & $S$ & $S_p$\\
    \hline
    Aq-A-5 &  2,316,893    & 634,793     & 712,232     & $2.294\times10^{6}$& 1    &$\times 1$\\
    Aq-A-4 & 18,535,972    & 634,793     & 5,715,467   & $2.868\times10^{5}$& 8    &$\times 8$\\
    Aq-A-3 & 148,285,000   & 20,035,279  &45,150,166   & $3.585\times10^{4}$& 64   &$\times 8$ \\
    Aq-A-2 & 531,570,000   & 75,296,170  &162,527,280  & $1.000\times10^{4}$& 229  &$\times 3.6$\\
    Aq-A-1 & 4,252,607,000 & 144,979,154 &1,306,256,871& $1.250\times10^{3}$& 1835 &$\times 8$\\
    \hline
  \end{tabular}
\end{table*}

\begin{table*}
  \caption{The number of \subhalos\ containing 20 or more particles and
    centres within a sphere of radius 250kpc/h from the fiducial
    centre found by each finder after standardised post-processing
    (see \S\ref{ssec:pipeline}). Three finders (\ahf, \rockstar\ \& \subfind) returned
    results from the highest resolution (level 1) within the timescale
    of this project. Below this we list the number of particles
    contained within the largest subhalo after post-processing.}
  \label{tbl:ResultsCent}
  \begin{tabular}{l *{11}{r}}
 \hline
         & \multicolumn{11}{c}{Number of \subhalos\ within 250kpc/h of
         the fiducial centre after post processing.}\\
    Name & \adaptahop&\ahf   &\hbt & \htd & \hsd &\hsf &\mendieta&\rockstar &\stf &\subfind&\voboz \\
    \hline 
    Aq-A-5 & 353      & 230       &    228 & 58  &136 &231    & 207     & 272      & 205 & 214    & 257 \\ 
    Aq-A-4 & 2497    & 1599    &  1544  &1265  &1075 & 1544  & 1493    & 1707     & 1521& 1433   & 1862 \\ 
    Aq-A-3 & -         & 11213     & 11693 &- &- & 11240 & 10948   & 11797    &10250& 10094  & 13343 \\ 
    Aq-A-2 & -         & 38441      & 39703 & -&- & 35445 & -       & 38489    & -   & 33135  & - \\ 
    Aq-A-1 & -         & 226802    & -        &- & -& - & -       & 235819   & -   & 221229 & - \\ 
    \hline
         & \multicolumn{11}{c}{Number of particles in the
         largest subhalo within 250kpc/h of
         the fiducial centre after post processing.}\\
    Aq-A-4 & 49076   & 77225 & 66470&69307&61581&73167&48387 & 78565     &56990& 50114  & 54685 \\ 
    \hline 
  \end{tabular}
\end{table*}

Both the halo finder catalogues (alongside the particle ID lists) and
our post-processing software (to be detailed below) are publically
available from the web site
\href{http://popia.ft.uam.es/SubhaloesGoingNotts}{http://popia.ft.uam.es/SubhaloesGoingNotts}
under the Tab ``Data''.

\section{The Comparison} \label{sec:Comparison}

We are going to primarily focus on comparing the location of \subhalos\
(both visually and quantitatively), the mass spectrum, and the
distribution of the peak value of the rotation curve. The comparison,
however, is based solely upon the provided particle lists and not the
subhalo catalogues as the latter are based upon each code's own
definitions and means to determine aforementioned properties and hence
possibly introducing ``noise'' into the comparison
\citep[cf.][]{knebe_haloes_2011}. In order to achieve a fair
comparison between the respective finders we produced a single
analysis pipeline which we used to post-process the particle lists
provided by each participating group. This ensured consistency across
our sample while removing differences due to the adoption of different
post-processing methodologies and the particular choice of threshold
criteria. The comparison detailed in this paper is restricted to this
uniform post-processed dataset. We intend to explore differences due
to different methodologies in a subsequent work. However, we stress at
the outset that our particular chosen post-processing methodology is
not intended to be unique nor do we put it forward as the {\it best}
way of defining a subhalo. Rather we use a single methodology so that
we can first answer the most fundamental question: if we agree on a
single subhalo definition do the different finders agree on the most
fundamental properties they recover?  Perhaps surprisingly we will see
that the answer to this question is broadly yes.

We did not consider in this paper efficiency of processing, as to make 
a fair comparison the codes would need to run on comparable machines
with a set amount of memory and processors. In this instance the finders
were run with the resources that were available to each of the participants.
Some indication of the capabilities of the respective finders may be deduced 
from Table~\ref{tbl:ResultsCent}.

\subsection{Post-processing pipeline}\label{ssec:pipeline}

Some finders (e.g. \ahf) include the mass (and particles) of a subhalo
within the encompassing host halo whereas others do not
(e.g. \subfind), preferring each particle to only be associated with a
single structure.  Either of these approaches has its pros and
cons. For instance, keeping the subhalo mass as part of the halo mass
makes it straightforward to calculate the enclosed dynamical mass of
any object. However, such an approach easily leads to multiple
counting of mass, particularly if there are many layers of the
substructure hierarchy.  In principle though it is not difficult to
transform from one definition to the other given knowledge of both the
halo and particle locations.  In our study, 5 of the 11 finders chose
to include the mass of \subhalos\ whereas the other 6 did
not. Following our principle of creating a uniform analysis pipeline
we processed all the particle lists to ensure that a particle could
only reside within a single structure.  To this end, we first sorted
the returned halo catalogue into mass order. Then starting from the
smallest halo we performed the centring, trimming and overdensity
checks detailed below to trim the subhalo uniformly. We then tagged
the particles contained within this object as being within a subhalo
before continuing to the next largest subhalo and repeating the
procedure ignoring particles already tagged as being used before. This
preserved the maximum depth of the subhalo hierarchy while ensuring
that a particle could only reside within a single subhalo. We should
remark that in practice excising all the sub-\subhalos\ from each
subhalo's particle list made little difference to any of the results
presented here as at any level of the subhalo hierarchy only around 10
per cent of the material is within a subhalo of the current halo. So
sub-\subhalos\ contribute only around 1 per cent of the halo mass,
although it can affect other properties such as the centre of mass.

All the particles belonging to the list each finder identified as
being associated with a subhalo were extracted from the original
simulation data files to retrieve each particle's position, velocity
and mass. From this data the centre of mass was first calculated,
before being refined based on consideration of only the innermost 10
per cent of these particles, sorted with respect to the initial centre
of mass. This procedure was repeated until a stable centre was found,
i.e. until the change in the position was below the actual force
resolution of the simulation. Once the centre had been defined the
particles were ordered radially from this point and a rotation curve
$GM(<r)/r$ and overdensity $M(<r)/(4\pi r^3/3)$ calculated until it
dropped below 200 times the critical density $\rho_{\rm crit}$
defining the subhalo radius $R_{200}$ and mass $M_{200}$. All
particles outside $R_{200}$ were removed which was essential in
particular for the phase-space finders who also considered already
stripped material as still being part of and belonging to the
subhalo. Please note that our post-processing pipeline does explicitly
not feature an unbinding procedure as this already formed part of
most halo finding algorithms. At this point the maximum circular
velocity, \vmax\ was obtained by smoothing the rotation curve and
locating its maximum by searching both inwards and outwards for a peak
in the rotation curve and taking the average of these two measures, a
process that stabilises the measure if the rotation curve is very flat
or noisy.

We emphasise that the precise subhalo properties are somewhat
sensitive to the definition of the halo centre. Various groups use the
centre-of-mass as the centre of all material enclosed within the
subhalo's radius (both with and without including substructure), the
centre-of-mass of some smaller subset (as here for example), the
location of the most bound particle, the location of the densest
particle or the minimum of the gravitational potential. 
Additionally different groups use different methodologies
for deciding whether or not a particle is bound to a halo as this
involves some decisions about the global potential and can be a very
time consuming process if done fully generally and
iteratively.

Finally the choice of where to place the subhalo edge is also
problematic. By definition the subhalo resides within some
in-homogeneous background density and so at some point particles cease
to belong to it and should rather be associated with the background
object. Different groups split the host halo from the subhalo in
different ways and there is no {\it correct} method. Without a uniform
choice these differences can swamp any differences due to actually
finding \subhalos\ or not. We stress that our post-processing (where
we treat each subhalo in isolation) can only remove particles from the
original list of those particles associated with a subhalo. We have
therefore tested whether or not our results are sensitive to our
choice of 200 as an overdensity parameter by re-running our analysis
with a tighter threshold of 500. Other than making all the subhalo
masses smaller this has no noticeable effect on the scatter of the
cumulative number counts. We therefore decided to stick to the
original choice of $R_{200}$ and $M_{200}$, respectively. Further,
throughout the subsequent comparison only haloes with more than 20
(bound) particles within \rth were used, although some finders
detected and returned haloes with less particles.

To summarise, our uniform post-processing pipeline involved the
following steps, applied iteratively where necessary:
\begin{itemize}
\item The subhalo catalogues were sorted into mass order.
\item Starting from the smallest subhalo, the particles associated with the
  current subhalo were obtained from the simulation data.
\item Only particles tagged as ``not used before'' were considered.
\item The centre-of-mass was iteratively calculated using the
  innermost 50 per cent of particles. (Originally we used the innermost 10 per cent
but found some of the more dispersed sub structures did not converge with this value).
\item A value for \rth\ was calculated based on an enclosed overdensity of 200
  times the critical density. 
\item The subhalo mass and rotation curve peak \vmax\ were computed based
  on particles inside \rth.
\item Only substructures containing more than 20 particles were retained.
\end{itemize}

\subsection{Visual comparison}

A visual representation of the location and size (based on \vmax) of the recovered
\subhalos\ at Aquarius level 4 from each of the finders is shown in
\Fig{fig:haloes}. A smoothed colour image of the underlying dark
matter density based on all particles from the original Aquarius data
are shown in one quadrant of the main halo, and this is overplotted with
the recovered \subhalos\ from each finder indicated by circles whose
size is scaled according to \vmax\ (specifically \vmax\ (in km/s)
divided by 3). This allows a visual comparison between the
finders. Only haloes with \vmax\ $> 10h^{-1}$km/s are shown. We
immediately see that most of the finders are very capable of
extracting the locations of the obvious overdensities in the
underlying dark matter field. Wherever you would expect to find a
subhalo (given the background density map) one is indeed
recovered. This demonstrates that substructure finders should be
expected to work well, recovering the vast majority of the
substructure visible to the eye. Additionally, if our aforementioned
post-processing is applied the quantitative agreement between the
finders is also excellent, with the extracted structures having very
similar properties between finders (see below). The majority of the
finders agree very well, reliably and consistently recovering nearly
all the \subhalos\ with maximum circular velocities above our
threshold.

While \Fig{fig:haloes} illustrates the agreement between the finders
at a single Aquarius level (in this case level 4, which all the
participating finders have completed), in \Fig{fig:plotcomp} we
construct a similar Figure to illustrate the agreement between
levels. We show the same quadrant at level 3 to level 1 for the three
finders that have completed the level 1 analysis (i.e., \ahf,
\rockstar\ \& \subfind); we deliberately omitted level 5 and level 4
as the former is not very informative and the latter has already been
presented in \Fig{fig:haloes}. As can be seen, the main difference
between the different levels is in the exact location of the
substructures. This changes because additional power was added to the
Aquarius initial power spectrum to produce the additional small
objects that form as the resolution is increased (fundamentally, the
Nyquist frequency has changed as there are more available tracers
within the higher resolution box). This extra power moves the
substructure around slightly, and these differences are amplified in
the, by definition, non-linear region of a collapsed object. Despite
this the ready agreement between the three finders at any single level
is clear to see and this is similarly true for both the other finders
(\hbt, \hsf) that completed level 2. We do not explore the effect of
changing the resolution on subhalo extraction in more detail here
because that is not the main point of this paper, which focuses on how
well different finders extract substructure relative to each
other. Also, this topic has already been well studied for \subfind\
using this same suite of models by \citep{springel_aquarius_2008}.

\subsection{Subhalo Mass Function}

\begin{figure*}
 \centering
 \includegraphics[width=1\linewidth]{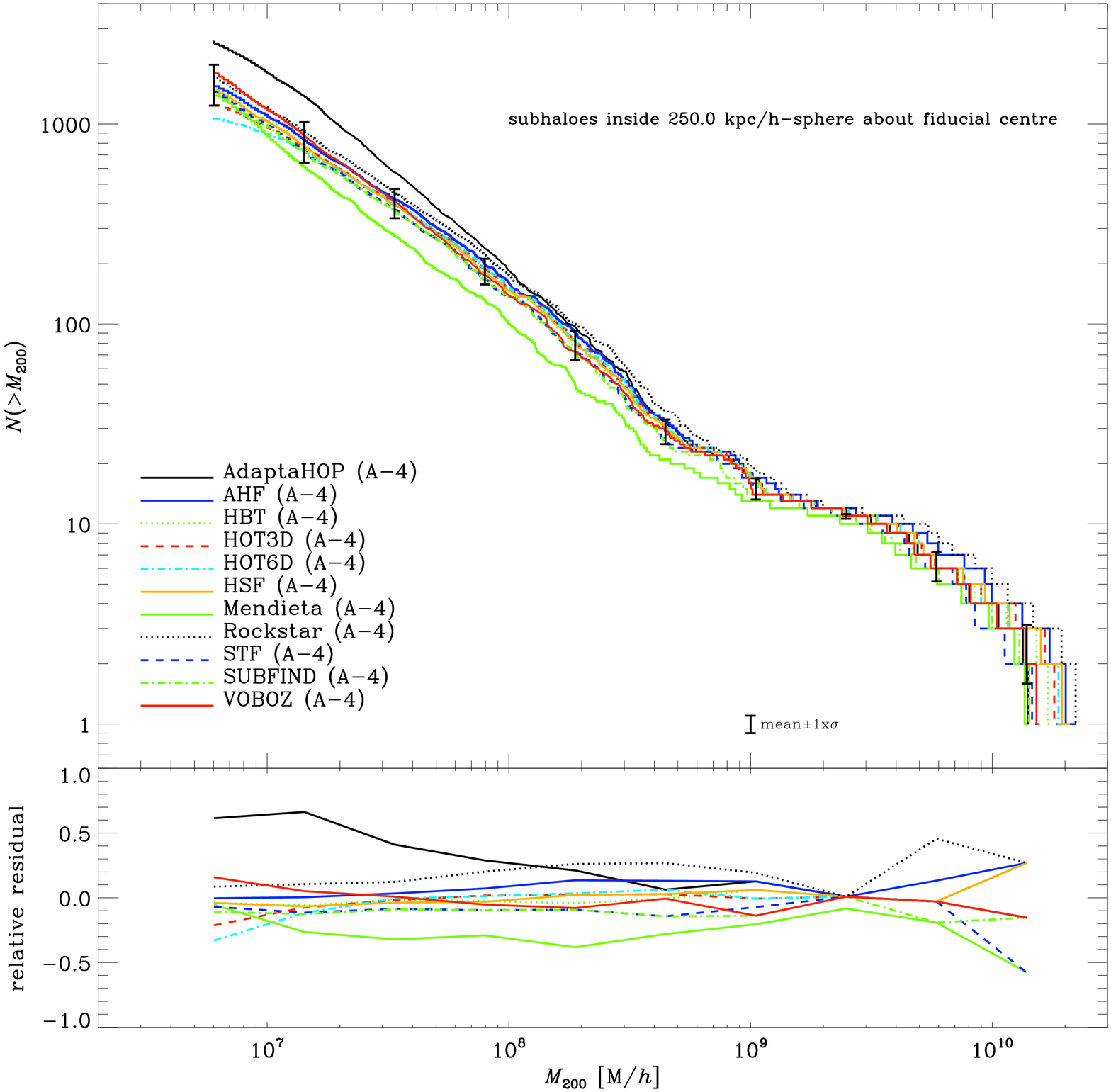} \\
\caption{Cumulative number count of \subhalos\ above the indicated mass
found ($M_{200}$) within a radius of 250 kpc/h from the fiducial halo centre after
standardised post-processing at resolution level 4 (see \S\ref{ssec:pipeline} for details).
The bottom plot shows the relative offset from the mean of the cumulative mass curve.}
\label{fig:massfunc}
\end{figure*}

\subsubsection{Level 4}
Perhaps the most straightforward quantitative comparison is simply to
count the number of \subhalos\ found above any given mass. For
Aquarius level 4 this produces the cumulative mass plot (based on
$M_{200}$) shown in \Fig{fig:massfunc}. Results from each
participating finder are shown as a line of the indicated
colour. Generally the agreement is good, with some intrinsic scatter
and a couple of outliers (particularly \adaptahop\ and \mendieta )
which do not appear to be working as well as the others, finding
systematically too many or too few \subhalos\ of any given mass
respectively. For \adaptahop\ we like to remind the reader that this
code does \textit{not} feature a procedure where gravitationally
unbound particles are removed; we therefore expect lower mass haloes
stemming from Poisson noise in the background host halo to end up in
the halo catalogue as well as haloes to have a higher mass in general
possibly explaining the distinct behaviour of this code. But typically
the scatter between codes is around the 10 per cent level except at
the high mass end where it is larger as each finder systematically
recovers larger or smaller masses in general. We like to remind the
reader again that this scatter is neither due to the
inclusion/exclusion of sub-\subhalos\ (which has been taken care of by
our post-processing pipeline) nor the definition of the halo edge: as
the 10 per cent differences still remain if choosing $R_{500}$ as the
subhalo edge.

\begin{figure*}
 \centering
  \includegraphics[width=1\linewidth]{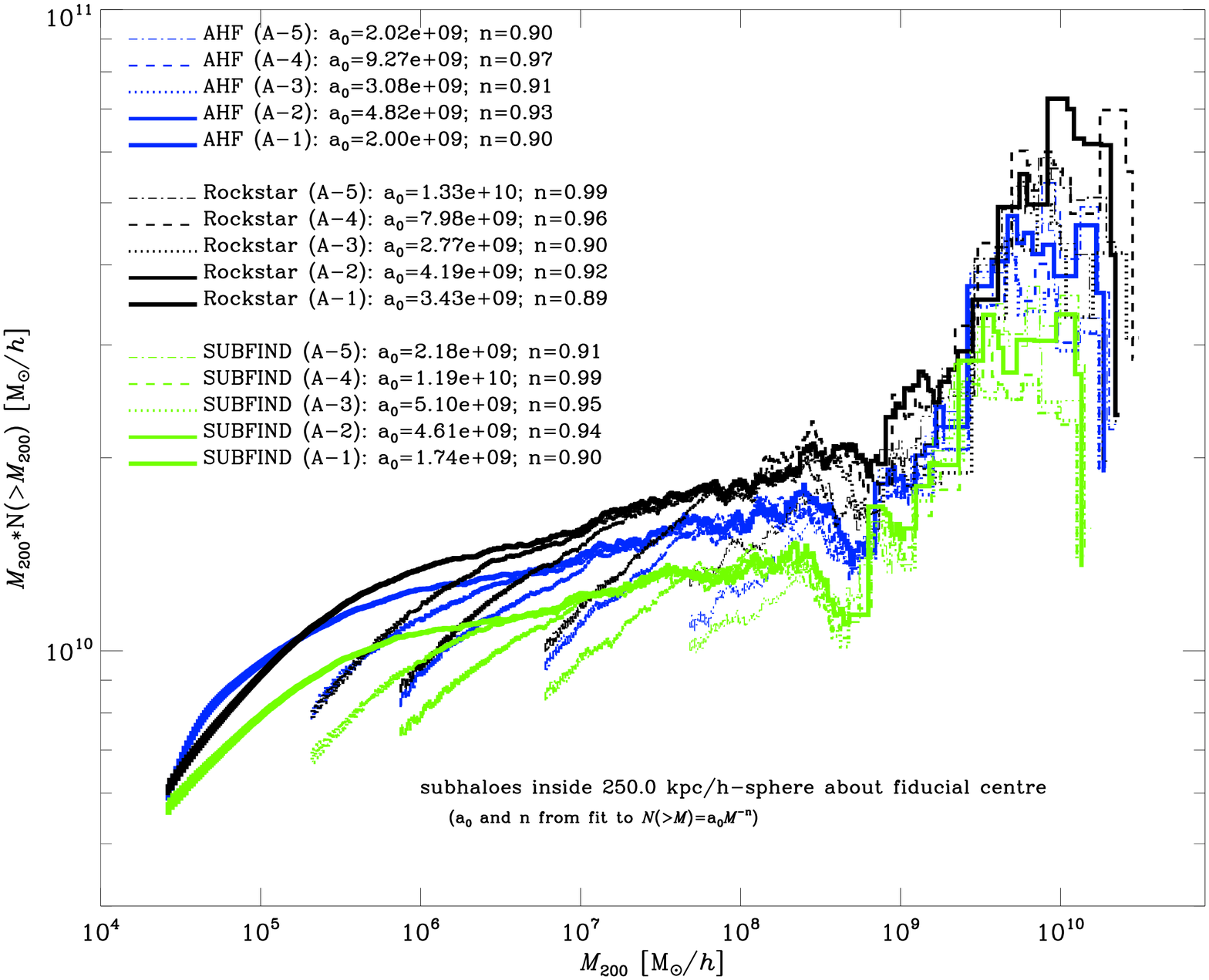}
  \caption{Cumulative subhalo mass function (multiplied by $M$ to
    compress the vertical dynamical range) for all five Aquarius
    levels for the \ahf, \rockstar, and \subfind\ finder. We fit the
    function $N(>M)/N_{\rm tot} = a_0 \times M^{-n}$ between the mass
    equivalent to 100 particles at each level and $10^9
    \mathrm{M_{\sun}/h}$. Note: the data has {\em not} been shifted for clarity
    but is as plotted.}
 \label{fig:fitAHF}
\end{figure*}

Table~\ref{tbl:ResultsCent} lists the number of \subhalos\ found that
contain 20 or more particles after the uniform post-processing
procedure detailed above had been performed and within $250$ kpc/h of
the fiducial centre of the main Aquarius halo at each level completed
for all the eleven finders that participated.  These number counts are
generally remarkably consistent, again with a few outliers as expected
from \Fig{fig:massfunc}. The majority of the finders are recovering
the substructures remarkably well and consistent, respectively.

As an additional quantitative comparison we list the number of
particles associated with the largest substructure found by each of
the finders as the last row of Table~\ref{tbl:ResultsCent}. All the finders
recover a structure containing $60,000$ particles $\pm 20$ per cent. As
\Fig{fig:massfunc} has shown there is a lot of residual scatter
for the highest mass haloes even when a uniform post-processing
pipeline is used. This is most likely due to the different unbinding
algorithms used in the initial creation of the substructure membership
lists which are particularly uncertain for these large structures. At
the other end of the substructure mass scale we have chosen to
truncate our comparison at \subhalos\ containing 20 particles as this
was shown to be the practical limit in \cite{knebe_haloes_2011}. Some
participants returned haloes smaller than this as this is their normal
practice. They all stress that such small \subhalos\ should be treated
with extreme caution but that there does appear to be a bound object
at these locations even if its size is uncertain. We have removed them
here for the purposes of a fair comparison.

Other plots that could be considered are those comparing the number 
of \subhalos\ against radial distance, or fractional mass against radial 
distance. Both these were produced and considered, but did not give any 
further insight to the comparison.

\subsubsection{All Levels}
Cumulative subhalo number counts like that shown for level 4 in
\Fig{fig:massfunc} can be calculated for all completed levels and
compared. As shown in \Fig{fig:plotcomp} while increasing the
resolution does not exactly reproduce the same substructures a
reasonable approximation is achieved and so we expect to find a set of
similar \subhalos\ containing more particles as we decrease the
individual particle mass between levels (i.e. any specific subhalo
should effectively be better resolved as the resolution increases). We
show the cumulative number counts for the finders \ahf, \rockstar, and
\subfind\ (multiplied by $M$ to compensate for the large vertical
scale) from level 5 to level 1 in \Fig{fig:fitAHF}. We show this as an
example and stress that similar plots with similar features could be
produced for any of the finders that completed level 2. The curve for
each level starts at 20 particles per halo and we like to stress that
no artificial shifting has been applied: any differences seen in the
plot are due to the different halo finding algorithms. Below about 100
particles per halo the cumulative number counts fall below the better
resolved curves, indicating that \subhalos\ containing between 20 and
100 particles are not fully resolved and should have a slightly higher
associated mass, also reported in \citet{muldrew_accuracy_2011}.
Above $10^9 \mathrm{M_{\sun}/h}$ the power law slope breaks as there
are less than 10 \subhalos\ more massive than this limit and the
number of these is a property of this particular host halo. For these
reasons we fit a power law of the form
\begin{equation}\label{eq:massfit}
  \frac{N(>M)}{N_{\rm tot}} = a_0 M^{-n}
\end{equation}
between 100 particles and $10^9 \mathrm{M_{\sun}/h}$ where the power
law breaks.  Here $a_0$ is a normalisation (capturing the rise in the
number of \subhalos\ due to the increase in resolution), $M$ is the
mass and $n$ is the power law slope. The fitted values of the
parameters by level are given in the legend for each finder. The
subhalo cumulative number count appears to be an unbroken power law --
at least in the range considered for the fitting. Similar results for
\subfind\ were found by \citet{springel_aquarius_2008}.

\begin{figure}
 \centering
  \includegraphics[width=1\linewidth]{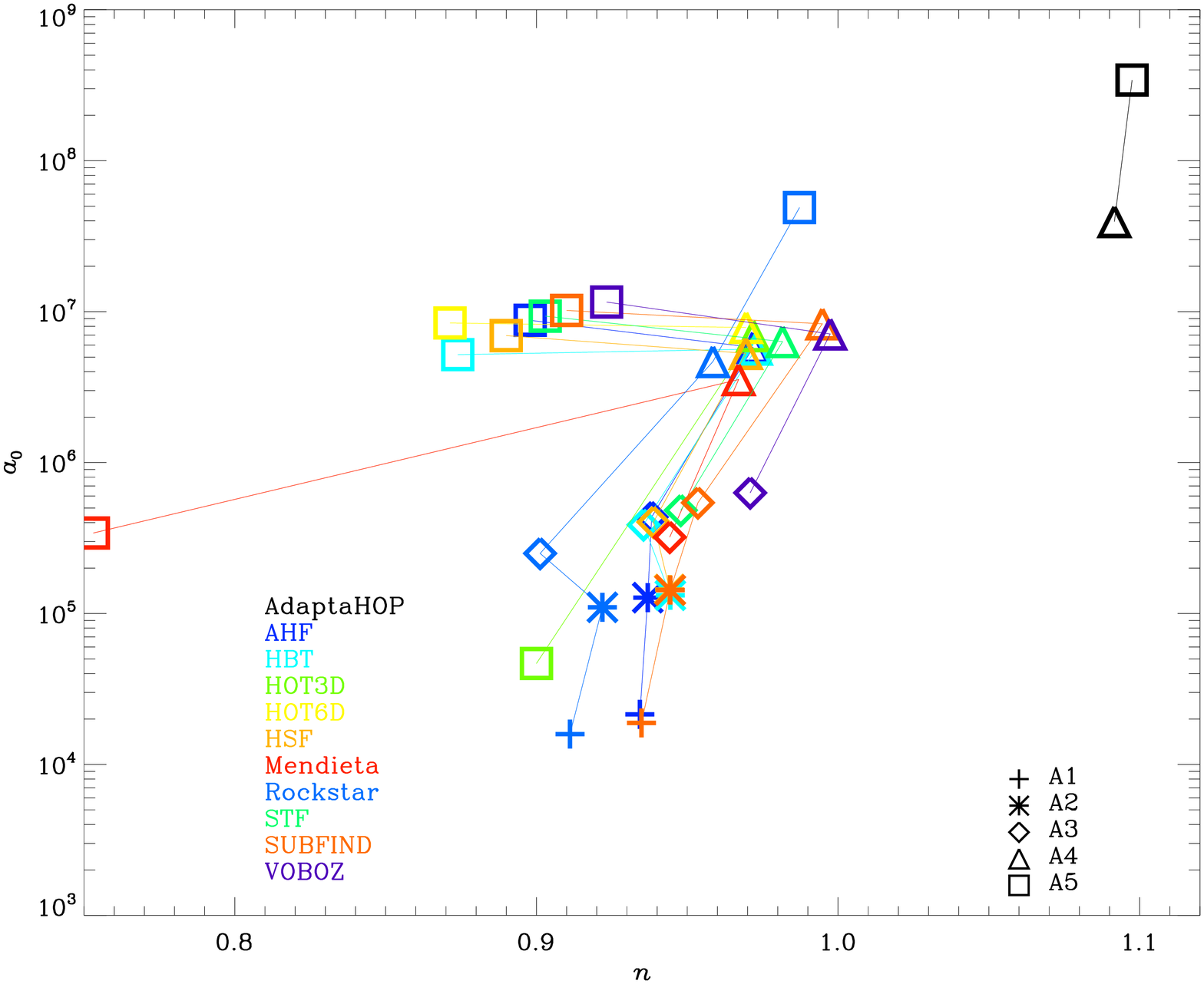}
  \caption{A comparison of the slope and normalisation of the fits
    of the mass function
    derived as per \Fig{fig:fitAHF} for all finders at all levels
    returned.}
 \label{fig:fitsubmass}
\end{figure}

We extended this particular analysis of fitting a single power-law to
the (cumulative) subhalo mass function to all finders at all available
levels and compare the values of $a_0$ and $n$ as a function of level
for all participating substructure finders in
\Fig{fig:fitsubmass}. There we find that at level 5 little can be said
because the fitting range is so narrow. At the lower, better resolved
levels good agreement is seen between the finders (clearly \adaptahop\
is a strong outlier on this plot, probably due to its lack of
unbinding as mentioned before when discussing \Fig{fig:massfunc}, and
\htd\ (as well as the first resolution step of \mendieta) is inverted
with respect to the main trend) and a consistent trend emerges: all
agree that the power law slope, $n$ is less than 1 and if anything
decreasing with increasing resolution.  Values of $n$ less than 1 are
significant because they imply that not all the mass is contained
within substructures, with some material being part of the background
halo. This has important ramifications for studies requiring the
fraction of material within substructures such as the dark matter
annihilation signal and lensing work. Although this result is robust
between all high-resolution finders we remind the reader that this is
for a single halo within a single cosmological model. However, it does
indicate that, as perhaps expected, the most important contribution to
substructure mass is from the most massive objects and that
progressively smaller structures contribute less and less to the
signal.

\subsection{Distribution of  \vmax}
If, instead of quantifying the total mass of each subhalo, we rather
use the maximum rotational velocity, \vmax\ to rank order the
\subhalos\ in size we obtain a generally much tighter relation (see
below). \citet{knebe_haloes_2011} already found that \vmax\ was a
particularly good metric for comparing haloes and we confirm this for
\subhalos. As \citet{muldrew_accuracy_2011} showed in figure 6 of
their paper, this is because for an NFW profile \citep{nfw_1997} the
maximum of the rotation curve is reached at less than 20 per cent of
the virial radius for objects in this mass range so \vmax\ is a
property that depends upon only the very inner part of the subhalo and
is not affected by any assumptions made about the outer edge. On the
other hand, it has also been shown
\citep{Ascasibar_and_Gottloeber_2008} that \vmax\ provides a
meaningful tracer of the depth of the gravitational potential
(i.e. the mass scale) of the halo.

\begin{figure*}
 \centering
 \includegraphics[width=1\linewidth]{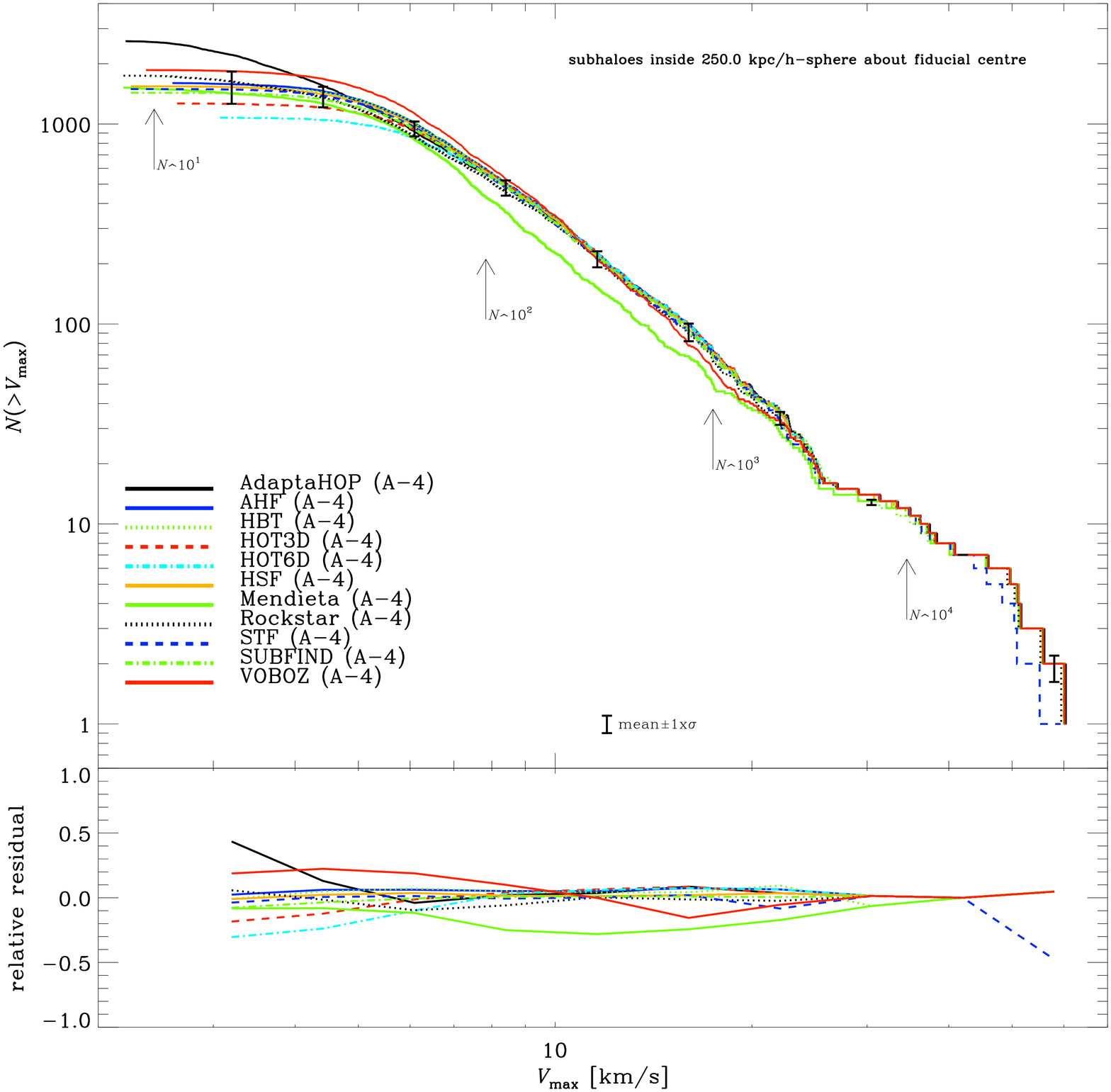}
\caption{Cumulative number count of \subhalos\ above the indicated \vmax\
value within a radius of 250 kpc/h from the fiducial halo centre after
standardised post-processing (see \S\ref{ssec:pipeline}). The arrows
indicate the number of particles interior to $r_{\rm max}$, the
position of the peak of the rotation curve. The bottom plot shows 
the relative offset from the mean of the cumulative count.}
\label{fig:vmaxplot}
\end{figure*}

\Fig{fig:vmaxplot} displays the cumulative \vmax\ for all the finders
for level 4 again. All the finders align incredibly well for the
largest \subhalos\ with \vmax $>$ 20 km/s. For \subhalos\ smaller than
this the alignment remains tighter than the total mass comparison down
to rotation velocities of around 6 km/s. At level 4 haloes of this
size contain around 80 particles in total, so \vmax\ is being
calculated from less than 20 particles at this point; the arrows give
an indication of the number of particles inside \Rmax.  \adaptahop,
despite its missing unbinding procedure, agrees well with other
finders for high rotation velocities as this particular statistic
probes inner regions of the \subhalos\ which are less affected by
unbound particles; and its deviation at the lower \vmax-end is due to
the existence of (small mass) fluke objects not removed by such an
unbinding step.

\subsection{Radial Mass Distribution}
\begin{figure*}
 \centering
 \includegraphics[width=1\linewidth]{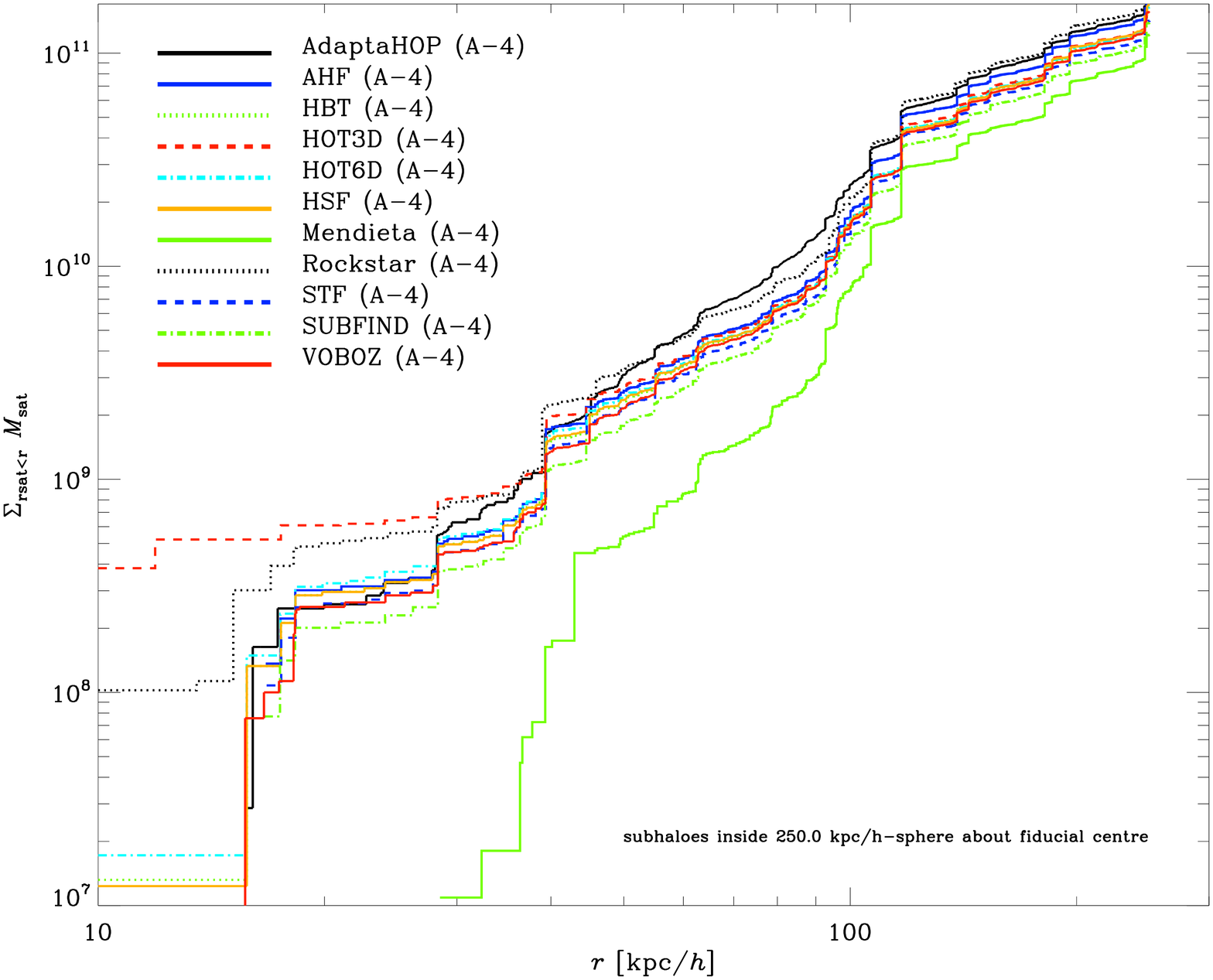}
\caption{Cumulative plot of the enclosed mass within \subhalos\ as a
  function of radial distance from the fiducial centre of the host
  halo.}
\label{fig:radialdist}
\end{figure*}

The accumulated total mass of material with \subhalos\ is measured by
ordering the subhalo centres in radial distance from the fiducial
centre of the halo and summing outwards,i.e. $\sum_{r_{\rm sat}<r}
M_{\rm sat}$. We include all post-processed \subhalos\ above our mass
threshold of 20 particles. As \Fig{fig:radialdist} demonstrates at
level 4 most of the finders (\ahf, \hbt, \hsd, \hsf, \stf, \voboz)
agree very well, finding very similar amounts of substructure both in
radial location and mass. \rockstar\ finds a little more structure,
particularly in the central region where its phase-space nature works
to its advantage and \subfind\ finds around a factor of 25 per cent less due
to its conservative subhalo mass assignment.

The \mendieta\ finder appears to show significantly different results to
the rest.  As previously noted the \adaptahop\ finder locates many
small \subhalos\ and these push up the total mass found in
substructure above that found by the others particular in the range
around 50-100kpc. We note that two of the three phase-space based
finders (\hsd, \& \hsf) have a radial performance
indistinguishable from real-spaced based finders. The only one to show
any difference is \rockstar\ and it remains unclear whether or not
this is in practice a significant improvement.

We further like to mention (though not explicitly shown here) that a
visual comparison akin to \Fig{fig:haloes} but
focusing on the central 20~\hkpc\ reveals that it is very likely that
the excess mass found in that inner region by some of the finders such 
as \htd\ may be due to mis-identifications of the host halo as \subhalos. 
In the very central region it is difficult for the underlying real-space Friends-of-Friends
methodology to distinguish structures from the background halo and
so can show up either as multiple detections, or no structure at all.

\section{Summary \& Conclusions} \label{sec:Summary}

We have used a suite of increasing resolution models of a single Milky
Way sized halo extracted from a self-consistent cosmological
simulation (i.e. the Aquarius suite \citep{springel_aquarius_2008}) to
study the accuracy of substructure recovery by a wide range of popular
substructure finders. Each participating group analysed independently
as many levels of the Aquarius-A dataset at redshift $z=0$ as they
could manage and returned lists of particles they associated with any
subhalo they found. These lists were post-processed by a single
uniform analysis pipeline. This pipeline employed a standard fixed
definition of the subhalo centre and subhalo mass, and employed a
standard methodology for deriving \vmax. This analysis was used to
produce cumulative number counts of the \subhalos\ and examine how
well each finder was able to locate substructure.

We find a remarkable agreement between the finders which are based on
widely different algorithms and concepts. The finders agree very well
on the presence and location of \subhalos\ and quantities that depend
on this or the inner part of the halo are amazingly well and reliably
recovered. We agree with \citet{knebe_haloes_2011} that \vmax\ is a
good parameter by which to rank order the haloes (in this case
\subhalos). However, we also show that as \vmax\ is only dependent
upon the inner 20 per cent or less of the subhalo particles around 100
particles are required to be within the subhalo for this measure to be
reliably recovered.  Quantities that depend on the outer parts of the
\subhalos, such as the total mass, are still recovered with a scatter
of around 10 per cent but are more dependent upon the exact algorithm
employed both for unbinding (intrinsic to each finder) and to define
the outer edge (given by the common post-processing applied here).

The most difficult region within which to resolve substructures is the
very centre of the halo which has, by definition, a very high
background density. In this region real-space based finders are
expected to struggle whereas the full six-dimensional phase-space
based finders should do better. In practice \rockstar\ is the only
phase-space based finder that shows any indication of this (and this
difference becomes {\it less} pronounced as the resolution is
increased); but we cannot rule out mis-identifications of the host
halo as \subhalos\ at this stage.  We conclude that, as yet, none of the
phase-space based finders present a significant improvement upon the
best of the more traditional real-space based finders.

Convergence studies indicate that identified \subhalos\ containing
less than 100 particles tend to be under-resolved and these objects
grow slightly in mass if a higher resolution study is used. This could
be due to the fact that particles in the outer regions of these
\subhalos\ are stripped more readily at lower resolution or it could
be an artefact of the difficulty of measuring the potential (and hence
completing any unbinding satisfactorily) with this small number of
particles. Several studies
\citep{kase_2007,pilipenko_2009,trenti_2010} have indicated
the 
unreliability of halo properties (other than physical presence) for
(sub)haloes of this size or less.

Fitting power law slopes to the convergence studies of each finder
indicates that the logarithmic slope of the cumulative number count is
less than 1. While this is only confirmed for a single halo within a
single cosmology, and ignoring any mass in tidal streams, the result
appears to be robust as it is found for all the high-resolution
finders employed in this study. This indicates that the larger
substructures are the most important ones and that higher levels of
the (sub)subhalo hierarchy play a less significant dynamical role.

We like to close with a brief note on the removal of gravitationally
unbound particles for \subhalos. We have seen that the omission of
such a procedure most certainly leads to rather distinct
results. However, we cannot convincingly deduce whether or not this
will lead to more small mass objects (as is the case for \adaptahop)
or to objects more massive in general (also seen for \adaptahop),
likely both will occur. But we confirm that the exact differences
between a subhalo catalogue based upon a halo finding method with and
without unbinding depend on the actual algorithm to collect the
initial set of particles to be considered part of the subhalo: we
performed an analysis of the level 4 data with \ahf\ switching the
unbinding part off ending up with a subhalo mass function that was
only different at the higher mass end (not shown here though) as
opposed to the \adaptahop\ results; but both these codes differ
substantially in the way of assigning the primary particle set to a
subhalo.

It should be noted that the Aquarius-A halo is a relatively quiescent halo \citep{wang_2011},
not having been subject to many mergers. Investigation of other haloes, and those
produced by other simulation code would be interesting to compare.
Therefore more studies focusing on the actual halo catalogues returned by each
finder (as opposed to the particle lists used here); other cosmological
simulations and different simulated scenarios (such as disrupted galaxies); 
the detailed analysis of sub-substructure (which is only really practical at level 1); 
and other subhalo properties such as spin parameter and shape, as well as more detailed
resolution studies for those codes providing an analysis of all levels
will be deferred to future papers.

\section*{Acknowledgements} \label{sec:Acknowledgements}

We wish to thank the Virgo Consortium for allowing the use of 
the Aquarius dataset and Adrian Jenkins for assisting with the data.

AK is supported by the {\it Spanish Ministerio de Ciencia e
Innovaci\'on} (MICINN) in Spain through the Ramon y Cajal programme
as well as the grants AYA 2009-13875-C03-02, AYA2009-12792-C03-03,
CSD2009-00064, and CAM S2009/ESP-1496. He further thanks Astrud
Gilberto for the shadow of your smile.

HL acknowledges a fellowship from the European Commission's Framework
Programme 7, through the Marie Curie Initial Training Network CosmoComp
(PITN-GA-2009-238356).

MN acknowledges support through Alex Szalay from the Gordon and Betty Moore Foundation.

YA is also supported by the Ramon y Cajal programme, as well as grant
AYA 2010-21887-C04-03. He would also like to warmly thank JO and
AK for their patience and their invaluable help while debugging the
post-processing routine of HOT+FiEstAS.

JXH is supported by the European Commissions Framework Programme
7, through the Marie Curie Initial Training Network Cosmo-Comp
(PITNGA-2009-238356) and partially supported by NSFC (10878001,11033006,
11121062) and the CAS/SAFEA International Partnership Program for
Creative Research Teams (KJCX2-YW-T23).  The calculations with HBT
were performed on the ICC Cosmology Machine, which is part of the DiRAC
Facility jointly funded by STFC, the Large Facilities Capital Fund of BIS,
and Durham University.

VS acknowledges partial support by SFB 881 `The Milky Way System' of the DFG.

PJE acknowledges financial support from the Chinese Academy of Sciences
(CAS), from NSFC grants (No. 11121062, 10878001,11033006), and by the
CAS/SAFEA International Partnership Program for Creative Research Teams
(KJCX2-YW-T23).

\bibliography{mn-jour,SubHaloes}

\begin{thebibliography}{}

\bibitem[\protect\citeauthoryear{{Angulo}, {Lacey}, {Baugh} \&
  {Frenk}}{{Angulo} et~al.}{2009}]{angulo_2009}
{Angulo} R.~E.,  {Lacey} C.~G.,  {Baugh} C.~M.,    {Frenk} C.~S.,  2009, MNRAS,
  399, 983

\bibitem[\protect\citeauthoryear{{Ascasibar}}{{Ascasibar}}{2010}]{ascasibar_20%
10}
{Ascasibar} Y.,  2010, Comput. Phys. Commun., 181, 1438

\bibitem[\protect\citeauthoryear{Ascasibar \& Binney}{Ascasibar \&
  Binney}{2005}]{ascasibar_2005}
Ascasibar Y.,  Binney J.,  2005, MNRAS, 356, 872

\bibitem[\protect\citeauthoryear{{Ascasibar} \& {Gottl{\"o}ber}}{{Ascasibar} \&
  {Gottl{\"o}ber}}{2008}]{Ascasibar_and_Gottloeber_2008}
{Ascasibar} Y.,  {Gottl{\"o}ber} S.,  2008, MNRAS, 386, 2022

\bibitem[\protect\citeauthoryear{{Behroozi}, {Wechsler} \& {Wu}}{{Behroozi}
  et~al.}{2011}]{behroozi_2011}
{Behroozi} P.~S.,  {Wechsler} R.~H.,    {Wu} H.-Y.,  2011, ArXiv preprint
  arXiv:1110.4372

\bibitem[\protect\citeauthoryear{{Boylan-Kolchin}, {Bullock} \&
  {Kaplinghat}}{{Boylan-Kolchin} et~al.}{2011a}]{boylan_mwlcdm_2011}
{Boylan-Kolchin} M.,  {Bullock} J.~S.,    {Kaplinghat} M.,  2011a, ArXiv
  preprint arXiv:1111.2048

\bibitem[\protect\citeauthoryear{{Boylan-Kolchin}, {Bullock} \&
  {Kaplinghat}}{{Boylan-Kolchin} et~al.}{2011b}]{boylan_toobig_2011}
{Boylan-Kolchin} M.,  {Bullock} J.~S.,    {Kaplinghat} M.,  2011b, MNRAS, 415,
  L40

\bibitem[\protect\citeauthoryear{{Boylan-Kolchin}, {Springel}, {White},
  {Jenkins} \& {Lemson}}{{Boylan-Kolchin} et~al.}{2009}]{boylan_mill2_2009}
{Boylan-Kolchin} M.,  {Springel} V.,  {White} S.~D.~M.,  {Jenkins} A.,
  {Lemson} G.,  2009, MNRAS, 398, 1150

\bibitem[\protect\citeauthoryear{{Contini}, {De Lucia} \& {Borgani}}{{Contini}
  et~al.}{2011}]{contini_2011}
{Contini} E.,  {De Lucia} G.,    {Borgani} S.,  2011, ArXiv preprint
  arXiv:1111.1911

\bibitem[\protect\citeauthoryear{{Cooper}, {Cole}, {Frenk}, {White}, {Helly},
  {Benson}, {De Lucia}, {Helmi}, {Jenkins}, {Navarro}, {Springel} \&
  {Wang}}{{Cooper} et~al.}{2010}]{cooper_2010}
{Cooper} A.~P.,  {Cole} S.,  {Frenk} C.~S.,  {White} S.~D.~M.,  {Helly} J.,
  {Benson} A.~J.,  {De Lucia} G.,  {Helmi} A.,  {Jenkins} A.,  {Navarro} J.~F.,
   {Springel} V.,    {Wang} J.,  2010, MNRAS, 406, 744

\bibitem[\protect\citeauthoryear{{De Lucia}, {Kauffmann}, {Springel}, {White},
  {Lanzoni}, {Stoehr}, {Tormen} \& {Yoshida}}{{De Lucia}
  et~al.}{2004}]{DeLucia_2004}
{De Lucia} G.,  {Kauffmann} G.,  {Springel} V.,  {White} S.~D.~M.,  {Lanzoni}
  B.,  {Stoehr} F.,  {Tormen} G.,    {Yoshida} N.,  2004, MNRAS, 348, 333

\bibitem[\protect\citeauthoryear{{di Cintio}, {Knebe}, {Libeskind}, {Yepes},
  {Gottl{\"o}ber} \& {Hoffman}}{{di Cintio} et~al.}{2011}]{dicintio_2011}
{di Cintio} A.,  {Knebe} A.,  {Libeskind} N.~I.,  {Yepes} G.,  {Gottl{\"o}ber}
  S.,    {Hoffman} Y.,  2011, MNRAS, 417, L74

\bibitem[\protect\citeauthoryear{{Diemand}, {Kuhlen}, {Madau}, {Zemp}, {Moore},
  {Potter} \& {Stadel}}{{Diemand} et~al.}{2008}]{diemand_vl2_2008}
{Diemand} J.,  {Kuhlen} M.,  {Madau} P.,  {Zemp} M.,  {Moore} B.,  {Potter} D.,
     {Stadel} J.,  2008, Nat, 454, 735

\bibitem[\protect\citeauthoryear{Elahi, Thacker \& Widrow}{Elahi
  et~al.}{2011}]{elahi_peaks_2011}
Elahi P.~J.,  Thacker R.~J.,    Widrow L.~M.,  2011, MNRAS, 418, 320

\bibitem[\protect\citeauthoryear{{Ferrero}, {Abadi}, {Navarro}, {Sales} \&
  {Gurovich}}{{Ferrero} et~al.}{2011}]{ferrero_2011}
{Ferrero} I.,  {Abadi} M.~G.,  {Navarro} J.~F.,  {Sales} L.~V.,    {Gurovich}
  S.,  2011, ArXiv preprint arXiv:1111.6609

\bibitem[\protect\citeauthoryear{{Freeman} \& {Bland-Hawthorn}}{{Freeman} \&
  {Bland-Hawthorn}}{2002}]{freeman_2002}
{Freeman} K.,  {Bland-Hawthorn} J.,  2002, ARA\&A, 40, 487

\bibitem[\protect\citeauthoryear{{Gao}, {White}, {Jenkins}, {Stoehr} \&
  {Springel}}{{Gao} et~al.}{2004}]{gao_2004}
{Gao} L.,  {White} S.~D.~M.,  {Jenkins} A.,  {Stoehr} F.,    {Springel} V.,
  2004, MNRAS, 355, 819

\bibitem[\protect\citeauthoryear{Gill, Knebe \& Gibson}{Gill
  et~al.}{2004}]{gill_evolution_2004}
Gill S.~P.,  Knebe A.,    Gibson B.~K.,  2004, MNRAS, 351, 399

\bibitem[\protect\citeauthoryear{{Gill}, {Knebe}, {Gibson} \& {Dopita}}{{Gill}
  et~al.}{2004}]{gill_2004b}
{Gill} S.~P.~D.,  {Knebe} A.,  {Gibson} B.~K.,    {Dopita} M.~A.,  2004, MNRAS,
  351, 410

\bibitem[\protect\citeauthoryear{Han, Jing, Wang \& Wang}{Han
  et~al.}{2011}]{han_resolving_2011}
Han J.,  Jing Y.~P.,  Wang H.,    Wang W.,  2011, Arxiv preprint
  arXiv:1103.2099

\bibitem[\protect\citeauthoryear{Hernquist}{Hernquist}{1990}]{Hernquist_1990}
Hernquist L.,  1990, ApJ, 356, 359

\bibitem[\protect\citeauthoryear{Jarosik, Bennett, Dunkley, Gold, Greason,
  Halpern, Hill, Hinshaw, Kogut, Komatsu, Larson, Limon, Meyer, Nolta, Odegard,
  Page, Smith, Spergel, Tucker, Weiland, Wollack \& Wright}{Jarosik
  et~al.}{2011}]{wmap_2011}
Jarosik N.,  Bennett C.~L.,  Dunkley J.,  Gold B.,  Greason M.~R.,  Halpern M.,
   Hill R.~S.,  Hinshaw G.,  Kogut A.,  Komatsu E.,  Larson D.,  Limon M.,
  Meyer S.~S.,  Nolta M.~R.,  Odegard N.,  Page L.,  Smith K.~M.,  Spergel D.,
  Tucker G.~S.,  Weiland J.~L.,  Wollack E.,    Wright E.~L.,  2011, ApJS, 192,
  14

\bibitem[\protect\citeauthoryear{{Kase}, {Makino} \& {Funato}}{{Kase}
  et~al.}{2007}]{kase_2007}
{Kase} H.,  {Makino} J.,    {Funato} Y.,  2007, PASJ, 59, 1071

\bibitem[\protect\citeauthoryear{{Klypin}, {Gottl{\"o}ber}, {Kravtsov} \&
  {Khokhlov}}{{Klypin} et~al.}{1999}]{klypin_overmerging_1999}
{Klypin} A.,  {Gottl{\"o}ber} S.,  {Kravtsov} A.~V.,    {Khokhlov} A.~M.,
  1999, ApJ, 516, 530

\bibitem[\protect\citeauthoryear{Klypin, Kravtsov, Valenzuela \& Prada}{Klypin
  et~al.}{1999}]{klypin_1999}
Klypin A.,  Kravtsov A.~V.,  Valenzuela O.,    Prada F.,  1999, ApJ, 522, 82

\bibitem[\protect\citeauthoryear{{Klypin}, {Trujillo-Gomez} \&
  {Primack}}{{Klypin} et~al.}{2011}]{klypin_bolshoi_2011}
{Klypin} A.~A.,  {Trujillo-Gomez} S.,    {Primack} J.,  2011, ApJ, 740, 102

\bibitem[\protect\citeauthoryear{{Knebe}, {Gill}, {Kawata} \& {Gibson}}{{Knebe}
  et~al.}{2005}]{knebe_2005}
{Knebe} A.,  {Gill} S.~P.~D.,  {Kawata} D.,    {Gibson} B.~K.,  2005, MNRAS,
  357, L35

\bibitem[\protect\citeauthoryear{Knebe, Knollmann, Muldrew, Pearce,
  {Aragon-Calvo}, Ascasibar, Behroozi, Ceverino, Colombi \& Diemand}{Knebe
  et~al.}{2011}]{knebe_haloes_2011}
Knebe A.,  Knollmann S.~R.,  Muldrew S.~I.,  Pearce F.~R.,  {Aragon-Calvo}
  M.~A.,  Ascasibar Y.,  Behroozi P.~S.,  Ceverino D.,  Colombi S.,    Diemand
  J.,  2011, MNRAS, pp 819--

\bibitem[\protect\citeauthoryear{Knollmann \& Knebe}{Knollmann \&
  Knebe}{2009}]{knollmann_ahf:_2009}
Knollmann S.~R.,  Knebe A.,  2009, ApJS, 182, 608

\bibitem[\protect\citeauthoryear{Kuhlen, Diemand \& Madau}{Kuhlen
  et~al.}{2008}]{kuhlen2008dark}
Kuhlen M.,  Diemand J.,    Madau P.,  2008, ApJ, 686, 262

\bibitem[\protect\citeauthoryear{{Libeskind}, {Knebe}, {Hoffman},
  {Gottl{\"o}ber} \& {Yepes}}{{Libeskind} et~al.}{2011}]{libeskind_2011}
{Libeskind} N.~I.,  {Knebe} A.,  {Hoffman} Y.,  {Gottl{\"o}ber} S.,    {Yepes}
  G.,  2011, MNRAS, 418, 336

\bibitem[\protect\citeauthoryear{Maciejewski, Colombi, Springel, Alard \&
  Bouchet}{Maciejewski et~al.}{2009}]{maciejewski_phasespace_2009}
Maciejewski M.,  Colombi S.,  Springel V.,  Alard C.,    Bouchet F.~R.,  2009,
  MNRAS, 396, 1329

\bibitem[\protect\citeauthoryear{{Maciejewski}, {Vogelsberger}, {White} \&
  {Springel}}{{Maciejewski} et~al.}{2011}]{maciejewski_2011}
{Maciejewski} M.,  {Vogelsberger} M.,  {White} S.~D.~M.,    {Springel} V.,
  2011, MNRAS, 415, 2475

\bibitem[\protect\citeauthoryear{Moore, Ghigna, Governato, Lake, Quinn, Stadel
  \& Tozzi}{Moore et~al.}{1999}]{moore_1999}
Moore B.,  Ghigna S.,  Governato F.,  Lake G.,  Quinn T.,  Stadel J.,    Tozzi
  P.,  1999, ApJ, 524, L19

\bibitem[\protect\citeauthoryear{Muldrew, Pearce \& Power}{Muldrew
  et~al.}{2011}]{muldrew_accuracy_2011}
Muldrew S.~I.,  Pearce F.~R.,    Power C.,  2011, MNRAS, 410, 2617

\bibitem[\protect\citeauthoryear{Navarro, Frenk \& White}{Navarro
  et~al.}{1997}]{nfw_1997}
Navarro J.~F.,  Frenk C.~S.,    White S. D.~M.,  1997, ApJ, 490, 493

\bibitem[\protect\citeauthoryear{Neyrinck, Gnedin \& Hamilton}{Neyrinck
  et~al.}{2005}]{neyrinck_voboz:_2005}
Neyrinck M.~C.,  Gnedin N.~Y.,    Hamilton A. J.~S.,  2005, MNRAS, 356, 1222

\bibitem[\protect\citeauthoryear{{Oh}, {Brook}, {Governato}, {Brinks}, {Mayer},
  {de Blok}, {Brooks} \& {Walter}}{{Oh} et~al.}{2011}]{Oh_2011}
{Oh} S.-H.,  {Brook} C.,  {Governato} F.,  {Brinks} E.,  {Mayer} L.,  {de Blok}
  W.~J.~G.,  {Brooks} A.,    {Walter} F.,  2011, ApJ, 142, 24

\bibitem[\protect\citeauthoryear{Pilipenko, Doroshkevich \&
  Gottl\"ober}{Pilipenko et~al.}{2009}]{pilipenko_2009}
Pilipenko S.,  Doroshkevich A.,    Gottl\"ober S.,  2009, Astronomy Reports,
  53, 976

\bibitem[\protect\citeauthoryear{{Pontzen} \& {Governato}}{{Pontzen} \&
  {Governato}}{2011}]{pontzen_2011}
{Pontzen} A.,  {Governato} F.,  2011, ArXiv preprint arXiv:1106.0499

\bibitem[\protect\citeauthoryear{Sgró, Ruiz \& Merchán}{Sgró
  et~al.}{2010}]{m._a._sgro_hierarchical_2010}
Sgró M.~A.,  Ruiz A.~N.,    Merchán M.~E.,  2010, BAAA, 53, 43

\bibitem[\protect\citeauthoryear{Springel}{Springel}{2005}]{springel_cosmologi%
cal_2005}
Springel V.,  2005, MNRAS, 364, 1105

\bibitem[\protect\citeauthoryear{Springel, Wang, Vogelsberger, Ludlow, Jenkins,
  Helmi, Navarro, Frenk \& White}{Springel
  et~al.}{2008}]{springel_aquarius_2008}
Springel V.,  Wang J.,  Vogelsberger M.,  Ludlow A.,  Jenkins A.,  Helmi A.,
  Navarro J.~F.,  Frenk C.~S.,    White S.~D.,  2008, MNRAS, 391, 1685

\bibitem[\protect\citeauthoryear{Springel, White, Frenk, Navarro, Jenkins,
  Vogelsberger, Wang, Ludlow \& Helmi}{Springel
  et~al.}{2008}]{springel_prospects_2008}
Springel V.,  White S. D.~M.,  Frenk C.~S.,  Navarro J.~F.,  Jenkins A.,
  Vogelsberger M.,  Wang J.,  Ludlow A.,    Helmi A.,  2008, Nat, 456, 73

\bibitem[\protect\citeauthoryear{{Springel}, {White}, {Jenkins}, {Frenk},
  {Yoshida}, {Gao}, {Navarro}, {Thacker}, {Croton}, {Helly}, {Peacock}, {Cole},
  {Thomas}, {Couchman}, {Evrard}, {Colberg} \& {Pearce}}{{Springel}
  et~al.}{2005}]{springel_millenium_2005}
{Springel} V.,  {White} S.~D.~M.,  {Jenkins} A.,  {Frenk} C.~S.,  {Yoshida} N.,
   {Gao} L.,  {Navarro} J.,  {Thacker} R.,  {Croton} D.,  {Helly} J.,
  {Peacock} J.~A.,  {Cole} S.,  {Thomas} P.,  {Couchman} H.,  {Evrard} A.,
  {Colberg} J.,    {Pearce} F.,  2005, Nat, 435, 629

\bibitem[\protect\citeauthoryear{Springel, White, Tormen \& Kauffmann}{Springel
  et~al.}{2001}]{subfind_2001}
Springel V.,  White S. D.~M.,  Tormen G.,    Kauffmann G.,  2001, MNRAS, 328,
  726

\bibitem[\protect\citeauthoryear{{Stadel}, {Potter}, {Moore}, {Diemand},
  {Madau}, {Zemp}, {Kuhlen} \& {Quilis}}{{Stadel}
  et~al.}{2009}]{stadel_ghalo_2009}
{Stadel} J.,  {Potter} D.,  {Moore} B.,  {Diemand} J.,  {Madau} P.,  {Zemp} M.,
   {Kuhlen} M.,    {Quilis} V.,  2009, MNRAS, 398, L21

\bibitem[\protect\citeauthoryear{{Trenti}, {Smith}, {Hallman}, {Skillman} \&
  {Shull}}{{Trenti} et~al.}{2010}]{trenti_2010}
{Trenti} M.,  {Smith} B.~D.,  {Hallman} E.~J.,  {Skillman} S.~W.,    {Shull}
  J.~M.,  2010, ApJ, 711, 1198

\bibitem[\protect\citeauthoryear{Vogelsberger, Helmi, Springel, White, Wang,
  Frenk, Jenkins, Ludlow \& Navarro}{Vogelsberger
  et~al.}{2009}]{vogelsberger2009phase}
Vogelsberger M.,  Helmi A.,  Springel V.,  White S.,  Wang J.,  Frenk C.,
  Jenkins A.,  Ludlow A.,    Navarro J.,  2009, MNRAS, 395, 797

\bibitem[\protect\citeauthoryear{{Wang}, {Navarro}, {Frenk}, {White},
  {Springel}, {Jenkins}, {Helmi}, {Ludlow} \& {Vogelsberger}}{{Wang}
  et~al.}{2011}]{wang_2011}
{Wang} J.,  {Navarro} J.~F.,  {Frenk} C.~S.,  {White} S.~D.~M.,  {Springel} V.,
   {Jenkins} A.,  {Helmi} A.,  {Ludlow} A.,    {Vogelsberger} M.,  2011, MNRAS,
  413, 1373

\bibitem[\protect\citeauthoryear{{Warnick}, {Knebe} \& {Power}}{{Warnick}
  et~al.}{2008}]{warnick_2008}
{Warnick} K.,  {Knebe} A.,    {Power} C.,  2008, MNRAS, 385, 1859

\bibitem[\protect\citeauthoryear{{White} \& {Rees}}{{White} \&
  {Rees}}{1978}]{white_1978}
{White} S.~D.~M.,  {Rees} M.~J.,  1978, MNRAS, 183, 341

\bibitem[\protect\citeauthoryear{Zavala, Springel \& Boylan-Kolchin}{Zavala
  et~al.}{2010}]{zavala_2010}
Zavala J.,  Springel V.,    Boylan-Kolchin M.,  2010, MNRAS, 405, 593

\end{thebibliography}
\bibliographystyle{mn2e} \label{sec:Bibliography}
\label{lastpage}

\end{document}